# Analysis of Sequence Polymorphism of LINEs and SINEs in Entamoeba histolytica

Submitted in partial fulfillment of the
requirements for the degree of

Master of Technology

in

Computational and Systems Biology

Submitted by
MOHAMMAD SULTAN ALAM

Under the guidance of
PROF. ALOK BHATTACHARYA & PROF. RAM RAMASWAMY

2010-2012

CENTRE FOR COMPUTATIONAL BIOLOGY & BIOINFORMATICS
SCHOOL OF COMPUTATIONAL & INTEGRATIVE SCIENCES

JAWAHARLAL NEHRU UNIVERSITY

NEW-DELHI 110067

# Acknowledgements

I take this opportunity to express a deep sense of gratitude towards my supervisors **Prof. Alok Bhattacharya** and **Prof. Ramakrishna Ramaswamy** for providing excellent guidance and encouragement throughout the project work. He introduced me to this area of work and was always there to help me in case of any need. Without his constant support and motivation this work could not have been completed successfully.

I would like to a acknowledge **Professor Sudha Bhattacharya** and **Dr. Vijay Pal Yadav**, School of Environmental Sciences, JNU, New Delhi, for providing data of *Entamoeba histolytica* SINE1s and helped me in understanding the details related to the retrotransposons.

My sincere thanks to Professor Karmeshu, Dean of School and also to Prof. Indira Ghosh, for providing us with all the help and ensuring that we do not face any problems in getting all the resources required for carrying out the research work.

I gratefully acknowledge my teachers, Dr. Narinder Singh Sahni, Dr. Pradipta Bandopadhyay, Dr. Naidu Subbarao, Dr. A. Krishnamachari, Dr. Andrew Lynn, Dr. Supratim Sengupta, Dr. Lovekesh Vig, Dr. Kushal Kumar Shah, Dr. Rashi Gupta, Prof. Rahul Roy, Dr. Devapriya Choudhary, for providing their valuable guidance throughout the course.

I also want to acknowledge my classmates (Navneet), who motivated and discussed various matters with me during my project.

I would also like to thank the staff of the school for their support and help.

Words fail me to express my appreciation to my God, family and friends (Tanveer Bhai) for their constant support and motivation which was indispensable for successful completion of the project.

I acknowledge the financial assistance and resources provided by D.B.T. and J.N.U.

Finally, I would like to thank everybody who contributed to the successful completion of the work in any way and express my apology to those whose names I could not mention.

Mohammad Sultan Alam

M.Tech II$^{nd}$ YEAR

SC & IS, JNU, New Delhi, India

*I would like to dedicate this thesis to my parents and family for their endless love, support and encouragement.*

# Table of Contents

# List of Figures

# List of Tables

# Chapter 1

# Introduction

*Entamoeba histolytica,* one of the most widespread and clinically important parasites that causes amoebiasis in most of the developing countries. According to one estimate approximately 50 million people worldwide develop amoebiasis with about 100,000 deaths per year (Ximenez, et al., 2009).

Transposable Elements (TEs) are nucleotide sequences able to jump from one position to another in the genome. TEs make up a large fraction of eukaryotic DNA, for examples 22% of the *Drosophila* (Kapitonov, et al., 2003), and 45% of human genome (Lander, et al., 2001).

Human genetic variations can cause phenotypic divergence as well as give an idea about evolutionary forces that shape genomes. Among genetic variations, single nucleotide polymorphisms (SNPs) have been widely studied, however, copy number variants, Indels (insertions/deletions) and TEs have also become useful for understanding genotype to phenotype relationship. Major class of TEs are Retrotransposons and these behave as endogenous mutagens since insertions can create genetic disorders.

LINEs (Long Interspersed Nuclear Elements) and SINEs (Short Interspersed Nuclear Elements) are retrotransposable elements. The term retrotransposable element is relevant to a broad class of nucleotide sequences whose insertion in the host DNA sequence is dependent on reverse transcription and therefore genetic information from RNA to DNA is directly transfered. In 1970, this idea was given by Temin and Baltimore when they identified and characterized retroviral reverse transcriptase for the first time. After that, nucleotide sequences coding for reverse transcriptase were detected not only in viruses (e.g. retroviruses, hepadnaviruses) but also in many eukaryotic transposons (e.g. retrotransposon of *E. histolytica*, bacterial retrointrons). Instability of genome increases due to retrotransposable elements, when copies of transcriptionally active sequences get inserted at new places during recombination.

Mutation allows for the prime variations on which selection, recombination, and genetic drift work (Drake, et al., 1999).

## 1.1 Viruses and parasites

**Virus: A** virus is an obligate intracellular parasite. Viruses can replicate only inside living host. Viruses are nonliving and have lack of mobility because they do not have capacity of independent growth and reproduction. For replication process, a virus must bind to or enter a living host cell, for example human or a bacteria cell, and during this process genetic material (RNA or DNA) is carried inside the cell. After this process the host cell replicates or produces numerous copies of the virus. Cells get killed or damaged when the replicated copies of the viruses leave host cells. In this way, released viruses infect new cells continuously and spread the disease. Viruses have different structures. There are two major components, genetic material (core) and protein coat or capsid (cover) that makes up all viruses. Many viral genomes have RNA as genetic material. In some viruses another covering is found that is called Envelope. Envelope is a membrane bilayer which is found outside of capsid. Viruses normally use host cell machinery and not only regulate their own reproduction but also control host cell processes, mostly for their own survival. Viruses can be host specific or tissue specific. For example, the human immunodeficiency virus (HIV) and polio virus attack the cells of the immune system and the nervous system respectively. Retroviruses are RNA viruses which have *gag*, *pol*, and *env* genes that, encode structural proteins, the Reverse Transcriptase (RT), and proteins embedded respectively in the viral coat. The reverse transcriptase plays a most important role in genome variations. Retroviruses makes DNA copy from RNA by using RT and this new copy of DNA is then used for the replication of viral genome (McClean, et al., 2004).

**Parasite:** Parasite is an organism that depends upon its host for shelter, food, and reproduction but does not provide anything to the host for its survival. If a parasite is present in our body, it consumes food we take for our nourishment.  Many species of fungi, protists and bacteria show parasitic nature. Protist parasites exist almost everyplace and it can enter inside us through different routes (insect bites, drinking dirty water, eating bare-assed fruits and vegetables). Many protist parasites are human and animal pathogens and these cause a range of diseases, varying in levels of morbidity and mortality. *Cryptosporidium, Giardia,* and *Entamoeba histolytica* are a few protist pathogens that commonly infect millions of people world wide. For example, amebiasis is caused by *E. histolytica* and nearly 40,000 people per annum die of this disease. *E. histolytica* display two stages, cysts and trophozoites. Only cysts can spread the disease while trophozoites actually cause the disease.

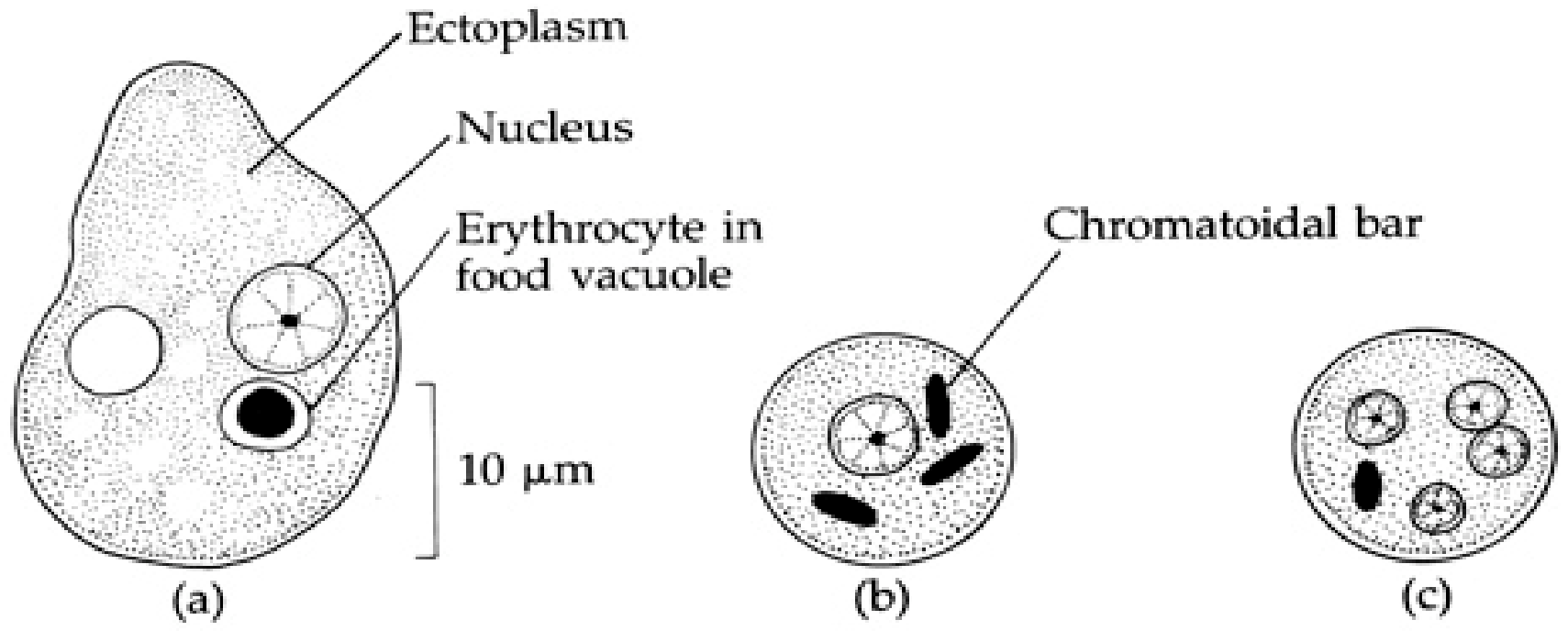

***Figure 1:*** *Entamoeba histolytica. (a) Trophozoite (b) Early cyst with four nuclei and rounded chomatoidal bars (c) Later cyst*

(Source: http://clem.mscd.edu/~churchcy/BIO3270/Images/Protozoans/E_histolytica.htm)

## 1.2 Genome Organization, Variation, Polymorphism

**Organization:** *E. histolytica* HM-1: IMSS is the first amoebazoan whose genome was sequenced first time in 2005 (Loftus, et al., 2005). According to the current annotation *E. histolytica* genome assembly is approximately 20.7 million base pairs (Mbp) in size (Lorenzi, et al., 2010). Table 1 shows a comparison of its genome with other organisms. (Lorenzi, et al., 2010; Clark, et al., 2007; Loftus, et al., 2005). The genome encodes 8,201 protein coding genes (Lorenzi, et al., 2010) covering ~ 49% of the total genome. Only 25% of genes have introns and only 6% have multiple introns (Clark, et al., 2007; Loftus, et al., 2005). The genome is highly A+T rich (67% in the coding region and 72% in the intergenic region). A study of 50651 codons used in *E. histolytica* genes proved a preference for NNU (43.1%) and NNA (42.4%) codons (Bhattacharya, et al., 2000).

**Table 1.** Genome summary statistics of *E. histolytica* and comparison with other sequenced genomes.

| Statistic / Organism | Genome size (Mbp) | G + C content (%) | Gene number | Av. gene size (bp) | Av. protein size (aa) | Gene density (kb per gene) | % Genes with introns | Av. intron size (bp) |
|---|---|---|---|---|---|---|---|---|
| ***E. histolytica*** [1] | 20.7 | 24.2 | 8201 | 1167 | 389 | 1.9 | 25.2 | 102 |
| ***P. falciparum*** [2] | 22.8 | 19.4 | 5268 | 2534 | 761 | 4.3 | 54 | 179 |
| ***D. discoideum*** [3] | 33.8 | 22.5 | 13541 | 1756 | 518 | 2.5 | 69 | 146 |
| ***S. cerevisiae*** [4] | 12.5 | 38 | 5538 | 1428 | 475 | 2.2 | 5 | ND |
| ***D. melanogaster*** [5] | 180 | 50 | 13776 | 1997 | 538 | 13.2 | 38 | ND |
| ***H. sapiens*** [6] | 2851 | 59 | 22287 | 27000 | 509 | 127.9 | 85 | 3365 |

Abbreviations: Mbp: million basepairs; kb: kilobase pairs; bp: basepairs; aa: amino acids, ND: not determined [1] (Lorenzi, et al., 2010; Clark, et al., 2007; Loftus, et al., 2005); [2] (Gardner, et al., 2002); [3] (Eichinger, et al., 2005); [4] (Goffeau, et al., 1996); [6] (Adams, et al., 2000); [8] (Venter, et al., 2001)

Currently more than 4000 viruses have been found that are distributed into 71 families. Viruses show all possible variations with respect to nature of their genomes, that is, single or double stranded DNA or single and double stranded RNA. RNA viruses are of small size in comparison with DNA viruses (Table 2).

**Table 2.** General characteristics of sequenced viral genomes.
The data was collected from NCBI, September 27, 2004 release from NCBI.

| **Viral class (no. of segments; range of the protein)**<br>**Examples** | **No. of Completed genomes** | **Size range (nt)** | **No. of proteins** |
|---|---|---|---|
| dsDNA (1; 5-698)<br>*Bovine polyomavirus*<br>*Ectocarpus siliculosus virus* | 414 | <br>4,697<br>335,593 | <br>6<br>240 |
| ssDNA (1,2, 8,10,11; 0-15)<br>*Coconut foliar decay virus*<br>*Milk vetch dwarf virus* | 230 | <br>1,360<br>10,958 | <br>6<br>11 |
| dsRNA (1,2,3,4,10,11,12; 1`-16)<br>*Mycovirus FusoV*<br>*Colorado tick fever virus* | 61 | <br>3,090<br>29,174 | 2<br>13 |
| ssRNA negative strand (1,2,3,4,6,8; 3-12)<br>*Borna disease virus*<br>*Rice grassy stunt virus* | 81 | <br>8,910<br>25,142 | 5<br>6 |
| ssRNA positive strand (1,2,3,4,5; 1-12)<br>*Ophiostoma novo-ulmi mitovirus 6-Ld*<br>*Murine hepatitis virus* | 421 | <br>2,343<br>31,357 | 1<br>11 |
| Satellites (1; 0-2)<br>*Rice yellow mottle virus satellite*<br>*Honeysuckle yellow vein mosaic virus-associated DNA beta* | 64 | <br>220<br>1,432 | 0<br>1 |

*(Source: http://www.ncbi.nlm.nih.gov/genomes/VIRUSES/viruses.html)*

**Variation:** Genome variation can be estimated on the basis of insertions and deletions (INDELs), genome rearrangements, Repeats, Gene duplication and Single Nucleotide

Polymorphisms (SNPs). A number of studies have been carried to estimate these variations and to understand molecular mechanisms that contribute to genome diversity and evolutionary advantages (Chandler, et al., 2001). Indels range from two to few thousands base pairs and a number of molecular processes including recombinational events and insertion sequence (IS) mediated events, expansion of repetitive sequences, or replication errors are responsible for it. These insertions and deletions have an important role in the study of evolution of genome size and fast changes in adaptability due of environment changes (Gregory, et al., 2003). For Instance, if we see indel spectrum of *Drosophila* and *Laupala* crickets, *Drosophila* has genome size 11 times smaller than *Laupala* but *in Laupala*, DNA loss is more than 40 times slower than that of *Drosophila*. Moreover, heterozygous indels is important for re-construction of DNA sequences and in the appraisal of ploidy and gnomic formation of hybrid organisms (Santos, et al., 2005). As agents, human-specific mobile elements insertions play an important role of molecular-genetic of anthropogenesis (Baskaev, et al., 2012). Each new insertion of mobile element provides the acceptor gene locus with the set of new functional sites for binding transcription factors that can make important changes in functioning of adjacent genes (Baskaev, et al., 2012). Retrotransposition is one of the main causes of gene expansion and creation of new genes. Expanded genes are very prone to mutations or variations leading to inactivation of genes and evolution of new genes or neo functionalization. Gene duplication and expansion plays an important role in enhancement of gene function and driving genetic variations (Kliebenstein, et al., 2008).

**Polymorphism:** Single-nucleotide polymorphism (**SNP**), is a DNA sequence variation that occur when a single nucleotide (A, T, C or G) in the genome sequence differs between members of a species or between paired chromosomes in an individual. This variation may occur due to replication error. SNPs, which create approximately 90% genetic variations in all human genome, are found after every 100 to 300 bases along the human genome. SNPs may fall within coding (gene) and non coding regions of the genome or intergenic regions between genes. In coding region, SNPs may or may not alter the amino acids sequences of proteins. SNP's that are in protein non-coding regions may change the specificities for gene splicing, factor binding, transcription or non-coding RNA sequence. These also help in genotype mapping and useful in understanding the function of commonly occurring phenotype

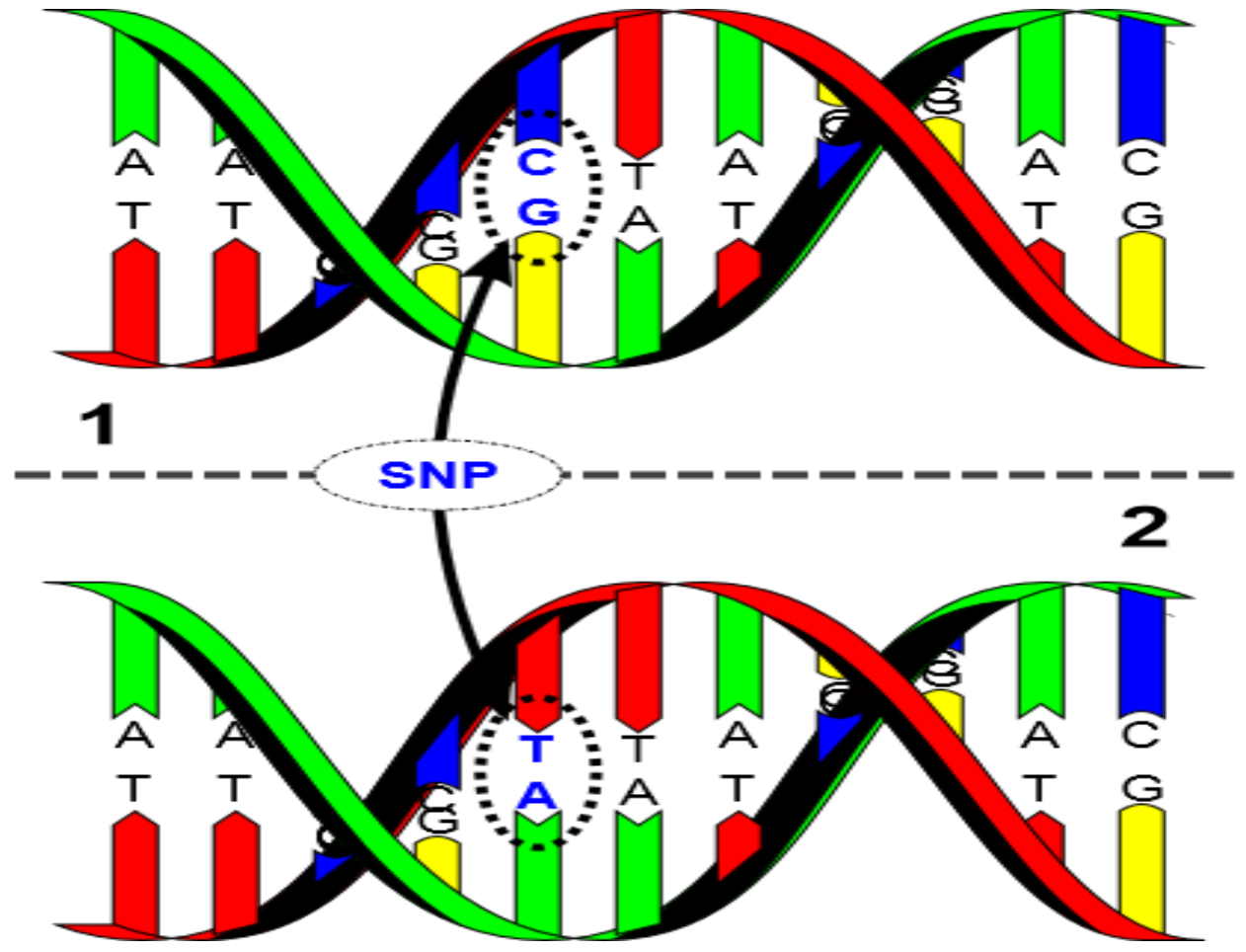

***Figure 2:** Single Nucleotide Polymorphism*

*(Source: http://en.wikipedia.org/wiki/File:Dna-SNP.svg)*

than the less common variants (Cargill, et al., 1999; Halushka, et al., 1999).

## 1.3 Repetitive DNA and Transposable Elements (TEs)

**Repetitive DNA:** Extent of repetitive DNA present in protist parasite vary to a larger extent including *E. histolytica* (Bhattacharya, et al., 2002; Wickstead, et al., 2003). It becomes easier to make accurate gene prediction and annotation if we have detailed identification of repeat elements in the genome. A large number of families of TEs have been described for *E. histolytica* genome. However, the most important repeat family is non-Long Terminal Repeat (LTR) retrotransposons. Within this family there are various subfamilies of LINE (Long Interspersed Nuclear Element) and SINE (Short Interspersed Nuclear Element) elements. LINE has subfamilies LINE1, LINE2, and LINE3 and SINE subfamilies are SINE1, SINE2, and SINE3. Also mutator related DNA transposon and one novel element ERE1 have been found in *E. histolytica* (Sharma, et al., 2001; Bakre, et al., 2005; Pritham, et al., 2005; Lorenzi, et al., 2008). Another element ERE2 has been reported in *E. histolytica* (Lorenzi, et al., 2008).

**Transposable elements (TEs):** TEs or mobile elements are sequences that are able to move within a genome or able to expand by making and inserting a number of their copies. TEs are commonly found in eukaryotic genome (Dombroski, et al., 1991) and are normally called jumping genes or Selfish genes (Doolittle, et al., 1980; Orgel, et al., 1980). The first TE was identified in maize by Barbara McClintock in 1948 (McClintock, et al., 1950). TEs have come out as diverse, copious, and omnipresent elements of not only eukaryotic genome, constituting up to 80% of nuclear DNA in plants, 3% to 20% in fungi, 3% to 52% in metazoans and approximately 45% in human genome but also in bacteria (insertion elements) (Wicker, et al., 2007). TEs can alter genome organization by element insertion or deletion and by homologous recombination between copies of elements. These events can alter nuclear architecture, stability of genome and gene regulation (Deininger, et al., 2003; Slotkin, et al., 2005). It is observed that most of the elements are found to be transcriptionally inactive but play an important role in size variation of homologous chromosomes. These are also thought to be involved in memory expansion and may be an expression of the evolutionary gains of genomic flexibility (Kidwell, et al., 2001).

There are two distinct classes of TEs on the basis of their structure and transposition: Class I transposons (retrotransposons) and Class II transposons (DNA transposons) (figure 3). Our work is focused on class I, so called retroelements or retrotransposons, which replicate through RNA copies. These use reverse transcriptase to make DNA copy from RNA strands.

Class I transposons move by “copy and paste” mechanism. Reverse transcription of RNA intermediate is needed for insertion of their cDNA at a position within the host genome. The retroelements can be classified in two major groups on the basis of the presence or the absence of long terminal repeats (LTRs). The first group is called LTR retroelements and the main representatives of this group are LTR retrotransposons, tyrosine recombinase retrotransposons and endogenous retroviruses. Long interspersed nuclear elements (LINEs) and short interspersed nuclear elements (SINEs) make up the second group, the so called non-LTR retroelements.

Non-LTR retrotransposons do not have tandem terminal and inverted repeats, instead these end mostly with a poly(A) tail at their 3` ends. LINE and SINE comprise ~11% of the 20.7 Mb genome of parasitic protist *E. histolytica* (Bakre, et al., 2005; Lorenzi, et al., 2010) making up a substantial fraction of the genome. LINE and SINE comprise ~11% of the 20.7 Mb genome of parasitic protist *E. histolytica* (Bakre, et al., 2005; Lorenzi, et al., 2010) making up a substantial fraction of the genome. In comparison to other protists E. histolytica is more abundant in terms of containing non LTR-retrotransposons. These are also present in most species of *Entamoeba* that have been studied suggesting their importance in genome evolution.

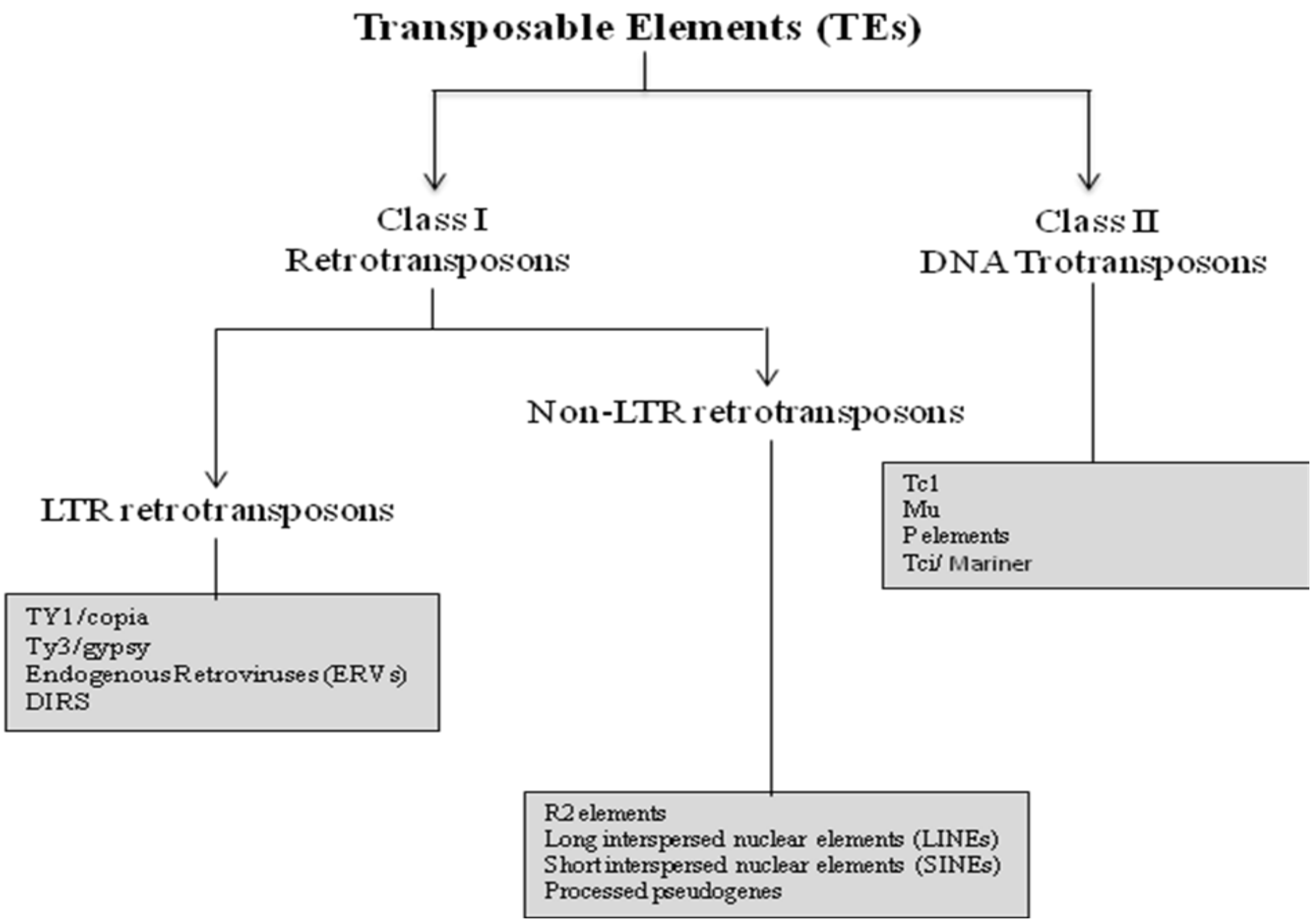

***Figure 3:*** *Classification of Transposable Elements (TEs)*

TEs are classified in two major classes; Class I- Retrotransposons and Class II- DNA transposons. The retrotransposons are further subdivided into two major groups; LTR (long terminal repeat) and non-LTR retrotransposons on the basis of presence or absence of LTR. Major members of each group are presented in gray boxes.

**Long Interspersed Nuclear Elements (LINEs):** These are autonomous non-LTR retrotransposons and generally widely found in eukaryotic genomes. About 17% of human DNA is constituted by these elements (Lander, et al., 2001). Most of the thousand copies of LINE are truncated and also carry large number of mutations so that, functionally only about 80-100 elements are potentially capable of retrotransposition (Brouha, et al., 2003). In human genome, full length LINE1 is 6 kb long. It has 900 nucleotide long 5` untranslated region (UTR) that works as an internal promoter for RNA polymerase II, a short 3` UTR, two open reading frames (ORF1 and ORF2), and a poly(A) tail. An endonuclease (EN) and reverse transcriptase (RT) are the common features in all non-LTR retrotransposons. The activities of these EN and RT are coded by ORF2. RTE element of *C.elegans* which lacks ORF1 domain and encodes only the endonuclease and reverse trancriptase, shows the minimal non-LTR retrotransposon (Figure 3). RT is highly conserved in all elements (Malik, et al., 1999). The endonuclease domain is of two types: apurinic/apyrimidinic (AP) endonuclease and type IIS restriction enzyme-like endonuclease (REL-ENDO). Till date, all reported non-LTR retrotransposons either have REL-ENDO or AP endonuclease.

**Short Interspersed Nuclear Elements (SINEs):** Apart from autonomous LINE elements, eukaryotic genomes are also constituted by non-autonomous, short (50-700 bp) non-coding sequences called SINEs. Like LINE, these elements contain an internal cleaved promoter, but SINEs are polymerase III transcribed elements and lack any ORF. In many SINE families, tRNA derived promoters have been described (Ohshima, et al., 1994) while some primate and rodent SINEs (e.g. Alu and B1 SINEs) contain 7SL gene derived promoters (Ullu, et al., 1984). Many SINE have sequence similarities at their 3` ends with their partner LINEs, for instance human Alu and L1 both terminate in poly(A) tail, flanked by target site duplication. Also there are many other examples that show marked sequence similarities at their 3` ends, such as LINEs and SINEs of the tortoise CR1LINE and Polymerase III (Ohshima, et al., 1996), the ruminant Bov-B LINE and Bob-tA SINE (Okada, et al., 1997), mammalian LINE2 and MIR SINE (Smit, et al., 1995) and eel UnaL2 LINE and UnaSINE1 (Kajikawa, et al., 2002). All these results suggest that SINEs use LINE machinery and the copying process starts from the similar 3` region (Kajikawa, et al., 2002; Ohshima, et al., 1996). For reverse transcription, SINE uses the reverse transcriptase of LINE elements.

**Class II transposons (DNA transposons):** Class II elements display simple transposition mechanism. DNA transposons move by “cut and paste” process i.e. the transposon is cut out of its position (like command / control x on our computer) and inserted into a new position (command / control v) of genome. This whole mechanism requires an enzyme transposase that is encoded within some of these transposons. Most DNA transposons have single open reading frame (ORF) and this ORF encode transposase. Transposons have inverted repeats on both ends. These inverted repeats are identical sequences reading in opposite directions. Some transposases need specific sequences as their target site and others can integrate at any place in the genome. The DNA at the target site is excised in an offset manner like the sticky ends.

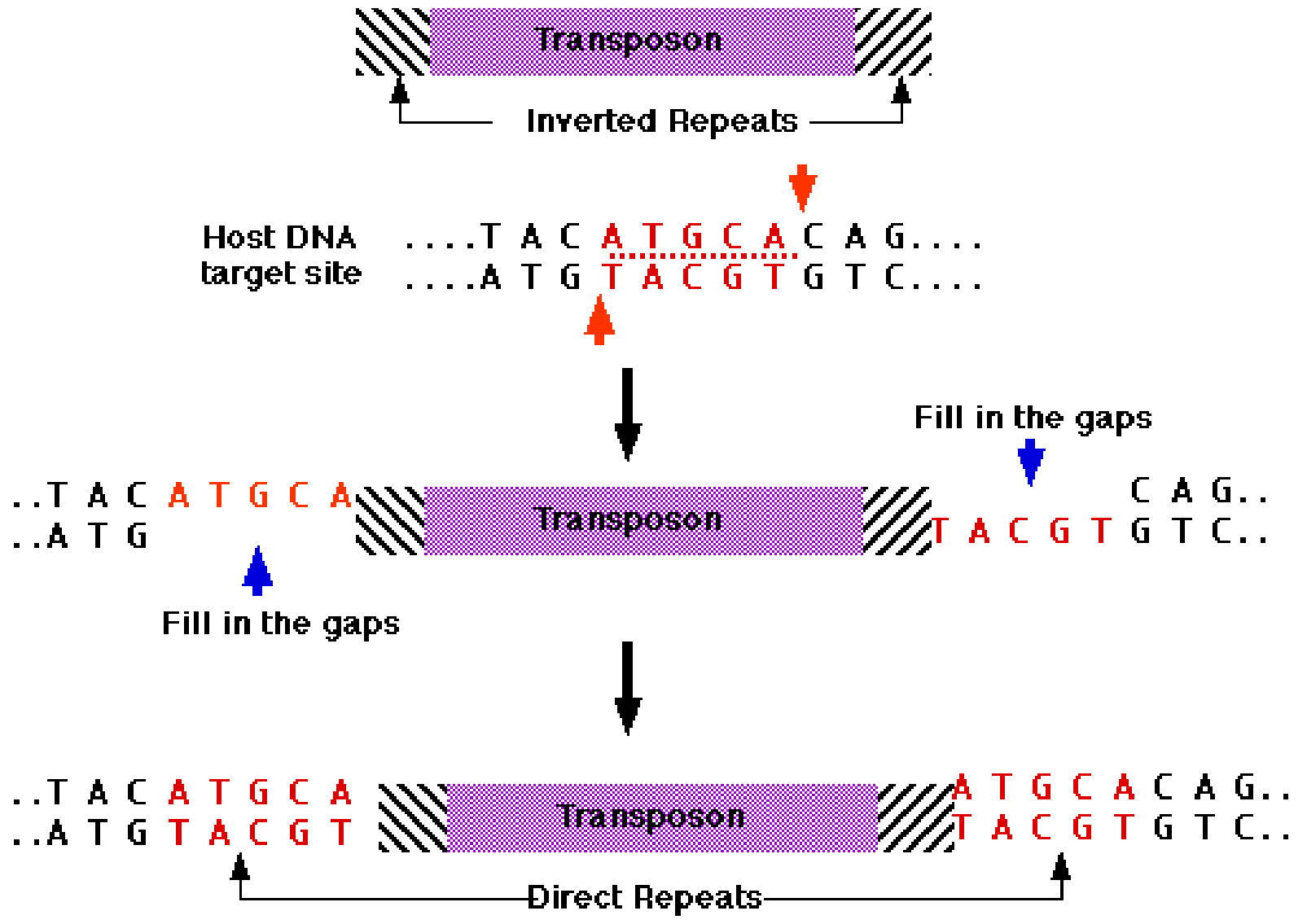

***Figure 4:** Mechanism of transposons elements*

Both the ends of transposon are inverted repeat. This transposon insertes into the host DNA and target site and then will create a gaps anad these gaps are filled by direct repeats.
*(Source: http://users.rcn.com/jkimball.ma.ultranet/BiologyPages/T/Transposons.html)*

It is not necessary that "cut and paste" mechanism will increase number of copies of transposons. If transposition occurs during host DNA replication then cut and paste mechanism may increase the copy number of transposon. Copy number can also increase if transposition takes place after replication of donor site before replication of the second site. DNA transposons seem to have lost the ability to transpose in the bulk of mammals. The last hierarchy-specific transposition occurred around 40 million year ago.

## 1.4 Mutation and Selection

Mutation is a change in a genomic sequence. Genomic sequence can be DNA sequence of a cell's genome or the DNA or RNA sequence of a virus. Mutation is an error that occurs during meiosis or DNA replication (Bertram, et al., 2000; Aminetzach, et al., 2005; Burrus, et al., 2004). For modelling of the genetic structure and evolution of populations, mutation rate is a main parameter. RNA viruses have high mutation rate in comparison to other organisms and this high mutation is responsible for the enormous adaptive capacity (Sanjua´n, et al., 2005). Mutation selection is a very old and fundamental concept in population genetics. The concept of natural selection is that, it increases the frequency of fit variants and the concept of mutations is that it introduces unfit variants. Due to these two effects, populations reach at equilibrium with balanced distribution of mutations. Mutation selection balance has been applied to explain the tenaciousness of undesirable genes, e.g. genetic diseases, and senescence. Consequences of mutation selection are realized by Manfred Eigen in 1971 when he studied this concept in long genomes (Eigen, et al., 1971). He determined that populations may not follow classic mutation selection balances in which the wildtype allele is most common. He found that population attain an equilibrium with a copious collection of mutant genotypes and a rare wildtype.

## 1.5 Quasispecies Theory

R. A. Fisher thought about species in a situation of optimal adaptability to their environment, hence they should be placed at one of the maxima of the adaptive landscape (Gavrilets, et al., 2004). Therefore a special sequence would be maximally adapted and the other sequences will have less replicative ability. For a single peak landscape form, there will be a distribution that is placed near the optimally adapted (master) sequence. In spite of the most adapted sequence, related to it (i.e. one, or several or many mutations away) there will be a group of less adapted sequences which exists together with the master sequence. It is exactly the high mutation rate impacting the viruses that tends to a heterogeneous ensemble and Eigen termed it as quasispecies. The concept of quasispecies has been applied in understanding sequence polymorphism encountered in viral populations and applied to explain an enormous array of evolutionary observations (Domingo, et al., 1992; Domingo, et al., 1997; Holland, et al., 1992; Steinhauer, et al., 1987). The original theory was given (Eigen, et al., 1971; Eigen, et al., 1979) to account for the variation of population evolution under the influence of mutation and selection. Initially this theory is used for studying pre-biotic evolution but in wider sense this theory also explains any population of reproducing organisms. Quasispecies theory is a concept of equilibrium (stationary) i.e. the final distribution of genotypes in population that have acquired to a stable state. In the quasispecies model, whatever initial condition of population we have, the final distribution of population will meet at equilibrium point due mutation and natural selection (Bull, et al., 2005). Also if the starting point of population is not at the equilibrium, then mutation and natural selection direct it toward equilibrium in the quasispecies model (Bull, et al., 2005). Suppose there are two genotypes $G_1$ and $G_2$.

Genotype $G_1$ has fitness $f_1$, and those $f_1$ offspring a fraction 1- $u_1$ retain the $G_1$ genotype. Its mutants are converted into genotype, $G_2$, which has less fitness $f_2$. $G_2$ produces its genotype with fidelity 1- $u_2$ and all of its mutants die. The product $f_1(1-u_1)$ is the number of $G_1$ offspring of an $G_1$ parent; refer as replacement rate of genotype $G_1$. Quasispecies theory has two novel properties, error threshold and error catastrophe (Bull, et al., 2005).

### 1.5.1 Error Threshold

A point at which the replacement rate of both genotype $G_1$ and $G_2$ are equal is called error threshold (Figure 5) (Bull, et al., 2005). $G_1$ has the higher replacement rate beyond this point, and $G_1$ is absent at all equilibria-the absence of back mutations ensures that it is not generated from $G_2$. An error threshold exists because there are deleterious mutations that affect some genotype more than others (Wilke, et al., 2005).

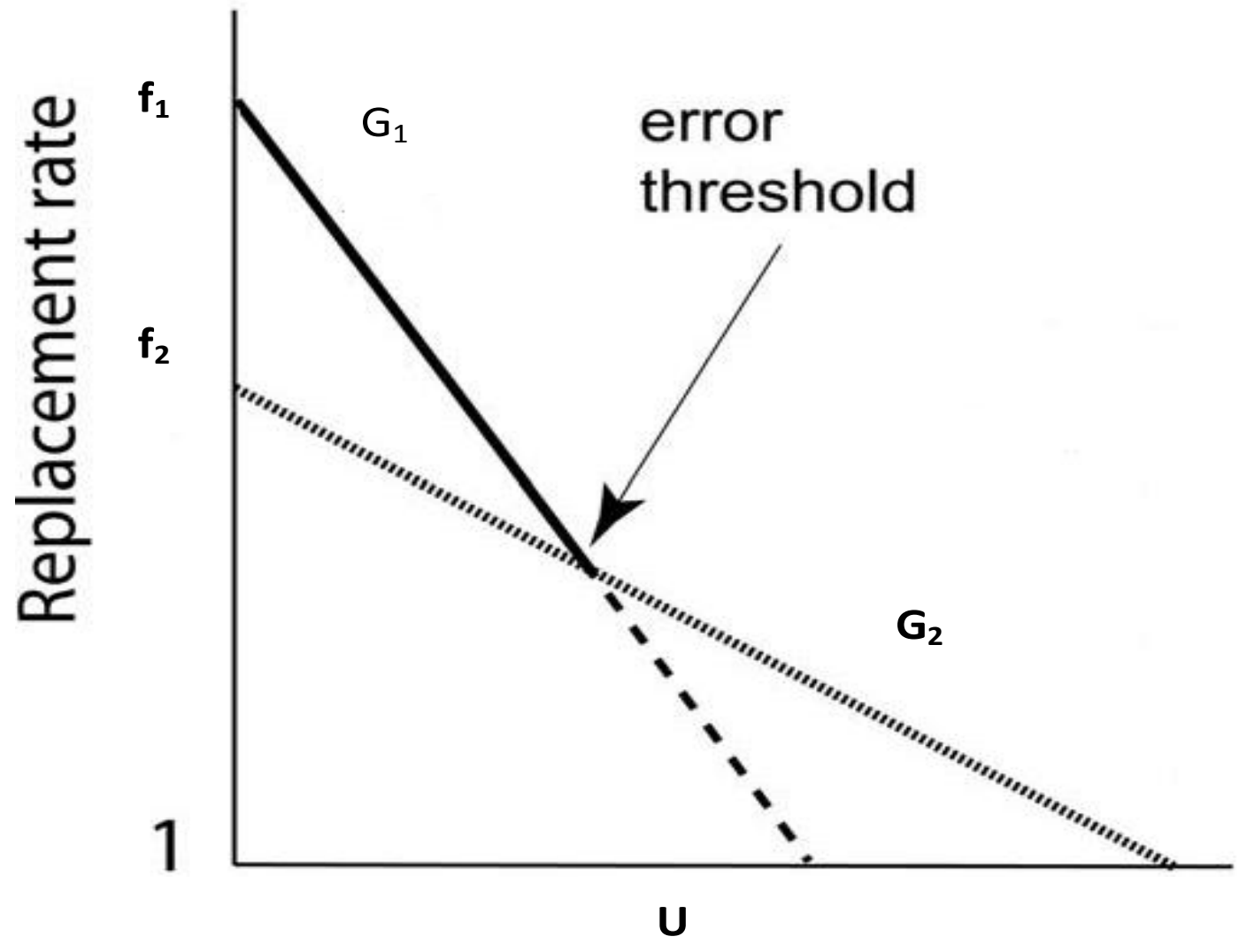

***Figure 5:*** *The Error Threshold*

The point at which the lines cross is the error threshold, beyond which $G_1$ is absent. (Adapted from Bull JJ et al., 2005)

### 1.5.2 Error Catastrophe

Error catastrophe is a property of mutation selection equilibrium and specific kind ofchanges in stationary point. Error catastrophe means complete loss of one favoured genotype $G_1$ at equilibrium point through deleterious mutations (Figure 6). The replacement rate functions of $G_1$ and $G_2$ cross at the error threshold giving rise to an error catastrophe i.e. complete loss of $G_1$ (see error threshold figure). Figure shows that the proportion of $G_1$ continuously decreases as mutation rate increases until the error threshold appeared (Bull, et al., 2005). Only $G_2$ will stay at higher mutation rates.

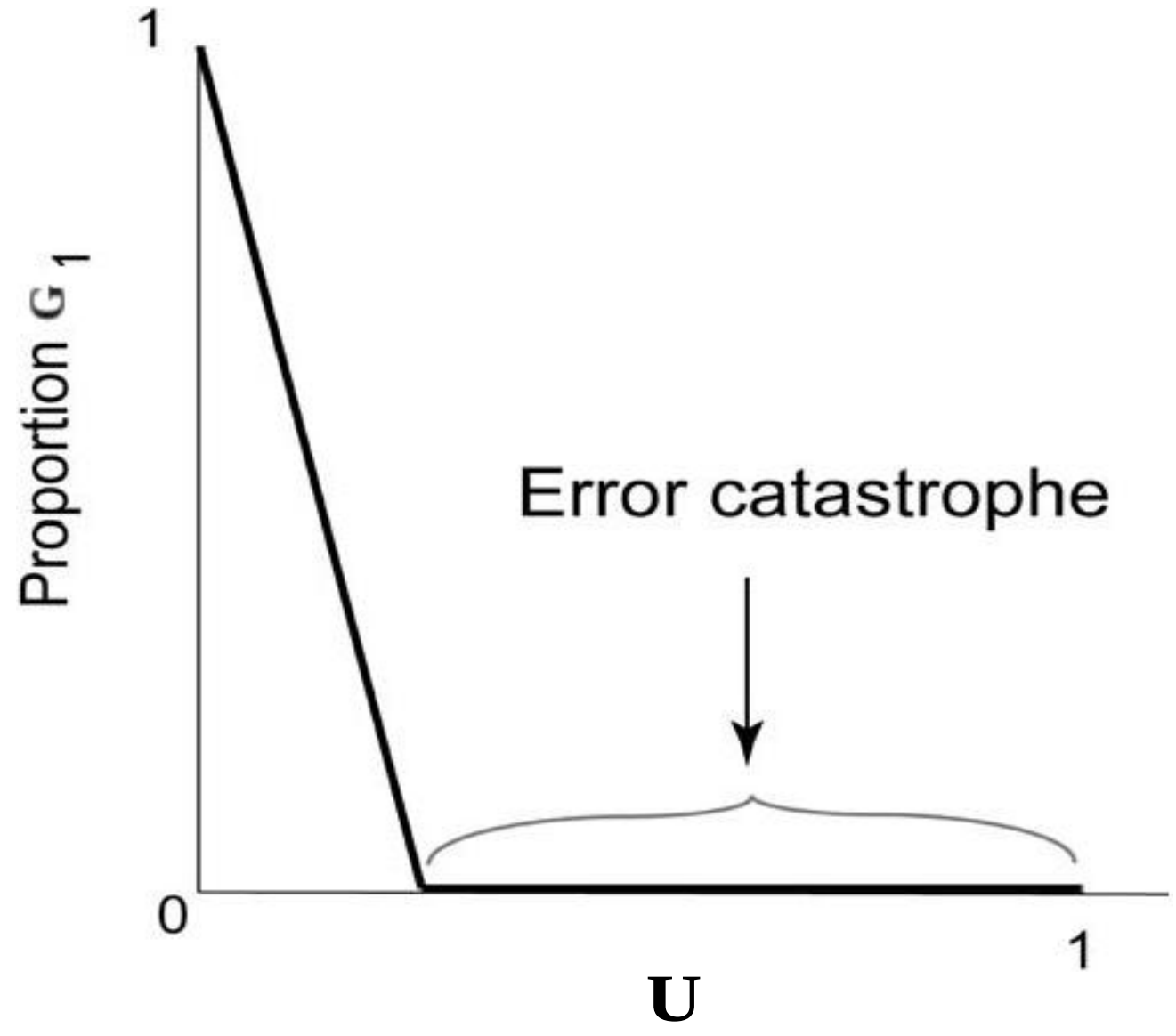

***Figure 6**: Error Catastrophe*

*(Adapted from Bull JJ et al., 2005)*

## 1.6 Aims and Objectives

Our aim is to study the sequence polymorphism in retrotransposable elements of *E. histolytica.* In this work, we have used computation approach to study the following:

The objectives of this dissertation are:

1. To find the mutation rate of retrotransposable elements of *E. histolytica*,
2. Analysis of the sequences variation within population and
3. Whether retrotransposable elements follow Quasispecies model or not.

# Chapter 2

# Methodology and Model

## 2.1 Datasets Used

In this study, two datasets have been used, first one being a published report (Huntley, et al., 2010) and the second has been obtained from Prof. Sudha Bhattacharya, School of Environmental Sciences, Jawaharlal Nehru University, New Delhi, India.

### 2.1.1 Dataset I (published) and Dataset II (Experimental)

**Data set I:** The genome sequence of *E. histolytica* having accession IDs AAFB00000000 with 1529 scaffolds was downloaded from NCBI database (http://www.ncbi.nlm.nih.gov/) in FASTA format. We extracted 393 EhSINE1s in various scaffolds using the coordinate of EhSINE1s of previous published work (Huntley, et al., 2010) by using custom Perl scripts. Out of these 393 EhSINE1s we have selected only 50 EhSINE1s for our analysis and the IDs and coordinates of these 50 sequences are mentioned in Table 3.

**Table 3.** Identity, Strand, Start Position, End Position and Lengths of *E. histolytica* SINE1s.

| SINE1 ID | Strand | Start Position (bp) | End Position (bp) | Length (bp) |
|---|---|---|---|---|
| **DS572593** | - | 697 | 150 | 548 |
| **DS571423** | - | 1030 | 483 | 548 |
| **DS571272** | + | 1423 | 1970 | 548 |
| **DS571151** | - | 154899 | 154352 | 548 |
| **DS571382** | - | 18914 | 18367 | 548 |
| **DS571347** | + | 2150 | 2696 | 547 |
| **DS571347** | + | 7763 | 8309 | 547 |
| **DS571317** | + | 27553 | 28099 | 547 |
| **DS571247** | - | 15177 | 14629 | 549 |
| **DS571179** | + | 37873 | 38420 | 548 |
| **DS571145** | + | 96823 | 97370 | 548 |
| **DS571278** | - | 6338 | 5791 | 548 |
| **DS571158** | + | 27154 | 27704 | 551 |
| **DS571257** | + | 45861 | 46408 | 548 |
| **DS571190** | + | 67757 | 68303 | 547 |
| **DS571494** | + | 2811 | 3357 | 547 |
| **DS571508** | - | 4500 | 3954 | 547 |
| **DS571346** | - | 26099 | 25553 | 547 |
| **DS571207** | - | 6989 | 6399 | 591 |
| **DS571661** | - | 3247 | 2657 | 591 |
| **DS571380** | - | 2811 | 2221 | 591 |
| **DS571159** | - | 21722 | 21132 | 591 |
| **DS571481** | - | 868 | 278 | 591 |
| **DS571828** | - | 3562 | 3038 | 525 |
| **DS571310** | - | 11751 | 11204 | 548 |
| **DS571255** | - | 5733 | 5183 | 551 |
| **DS571147** | - | 112278 | 111732 | 547 |
| **DS571487** | + | 12336 | 12883 | 548 |
| **DS572093** | - | 1734 | 1187 | 548 |
| **DS571277** | - | 39116 | 38571 | 546 |
| **DS571375** | + | 11554 | 12140 | 587 |
| **DS571499** | - | 10454 | 9868 | 587 |
| **DS571601** | - | 8125 | 7539 | 587 |
| **DS571333** | + | 24586 | 25172 | 587 |
| **DS572494** | + | 1 | 559 | 559 |
| **DS571897** | + | 627 | 1193 | 567 |
| **DS571442** | + | 14666 | 15230 | 565 |
| **DS571668** | + | 993 | 1556 | 564 |
| **DS571208** | - | 64860 | 64295 | 566 |
| **DS571874** | + | 2490 | 3045 | 556 |
| **DS571827** | - | 3250 | 2695 | 556 |
| **DS572564** | + | 309 | 865 | 557 |
| **DS572142** | + | 1214 | 1723 | 510 |
| **DS572182** | + | 493 | 964 | 472 |
| **DS571675** | - | 2982 | 2491 | 492 |
| **DS571660** | - | 757 | 270 | 488 |
| **DS572664** | - | 830 | 359 | 472 |
| **DS571250** | - | 22245 | 21673 | 573 |
| **DS571250** | - | 42224 | 41651 | 574 |
| **DS571151** | - | 120325 | 119752 | 574 |

**Data sets II:** We had also collected the experimental data of EhSINE1s from Dr. Vijay and Prof. Sudha Bhattacharya laboratory (personally), JNU, New Delhi (Vijay, et al., 2012).

### 2.1.2 Identification of best hits of EhSINEs

Similar sequences were identified using two kinds of alignment methods: global, which align whole sequence, and local, which align only highly similar subsequences. Needleman-Wunsch algorithm is used for global alignment and Smith-Waterman algorithm or BLAST is used for local alignment.

### 2.1.3 Local alignments for the EhSINEs

In order to identify SINE families, we selected one random sequence from 393 sequences and make it as a query sequence. The remaining 392 sequences were searched to find closest homologs and the process is repeated till all SINEs are classified. Homologs were defined by sequence similarity ≥ 98% and query coverage ≥ 80%. In this way, we have grouped all 393 sequences. We had selected those groups which contain atleast two best hits. In this way, we identified 8 groups from 50 sequences which satisfy our criteria and one group each containing 18, 6, 6, 5, 4, 4, 4 and 3 EhSINE1. Mismatches were identified from alignments. Example of blast results is as below:

################################################################

BLASTN 2.2.25+

Reference: Zheng Zhang, Scott Schwartz, Lukas Wagner, and Webb
Miller (2000), "A greedy algorithm for aligning DNA sequences", J
Comput Biol 2000; 7(1-2):203-14.

################################################################
================================================================

Database: d.fasta
392 sequences; 212,192 total letters

Query= gi|DS572593|SP-EP:697-150|548 bp

Length=548

> gi|DS571423|SP-EP:1030-483|548 bp
Length=548

Score = 990 bits (536), Expect = 0.0
Identities = 544/548 (99%), Gaps = 0/548 (0%)
Strand=Plus/Plus

================================================================

```
Query  1    AGATCGAAGGTGGCATGTCTGAAACACCACACATAAACCCTAGTACAAATTCATTCTTCG  60
            |||||||||||||||.||||||||||||||||||||||||||||||||||||||||||||
Sbjct  1    AGATCGAAGGTGGCACGTCTGAAACACCACACATAAACCCTAGTACAAATTCATTCTTCG  60

Query  61   ACTCTCCCAGTTATTATCTGGTTATGACGGTGCTTTTGAATTAGGAATGTATTAGGGAAT 120
            |||||||||||||||||||||||||||||||.||||||||||||||||||||||||||||
Sbjct  61   ACTCTCCCAGTTATTATCTGGTTATGACGGTCCTTTTGAATTAGGAATGTATTAGGGAAT 120

Query  121  GCTGCAAAGGGTGCAGCAAGAGAATACAGTAGAATATTACATGGATGTAATATAAGAATC 180
            ||||||||||||||||||||||||||||||||||||||||||||||||||||||||||||
Sbjct  121  GCTGCAAAGGGTGCAGCAAGAGAATACAGTAGAATATTACATGGATGTAATATAAGAATC 180

Query  181  TACTGAAGTGTGGGTATGACTAAAAGAAGATTAGTCAAAGTAAGACTAAAAAGAAGATTA 240
            ||||||||||||||||||||||||||||||||||||||||||||||||||||||||||||
Sbjct  181  TACTGAAGTGTGGGTATGACTAAAAGAAGATTAGTCAAAGTAAGACTAAAAAGAAGATTA 240

Query  241  GTCAAAGTAATACAGTAGTAATAAAATGATTCCTTCTTCCATTCAtaaaataagaaaaat 300
            |||||||||||||||||||||||||||||||||||||.||||||||||||||||||||||
Sbjct  241  GTCAAAGTAATACAGTAGTAATAAAATGATTCCTTCTCCCATTCATAAAATAAGAAAAAT 300
```

```
Query  301   gaaattccttaaaattaaggcagaaaacaaacaaaggcttaaaaagaagaaataagcaga 360
             ||||||||||||||||||||||||||||||||||||||||||||||||||||||||||||
Sbjct  301   GAAATTCCTTAAAATTAAGGCAGAAAACAAACAAAGGCTTAAAAAGAAGAAATAAGCAGA 360

Query  361   agaagtttgaaaaaccttaataggaagaaataaagcaaagaagTGCTTTCCTCATTTTGC 420
             |||.||||||||||||||||||||||||||||||||||||||||||||||||||||||||
Sbjct  361   AGAGGTTTGAAAAACCTTAATAGGAAGAAATAAAGCAAAGAAGTGCTTTCCTCATTTTGC 420

Query  421   AAGAAAAACATAAAGAATAGGTTTAACAAAGAGATTACTCTTTTTTAATAAGCTCAGGGA 480
             ||||||||||||||||||||||||||||||||||||||||||||||||||||||||||||
Sbjct  421   AAGAAAAACATAAAGAATAGGTTTAACAAAGAGATTACTCTTTTTTAATAAGCTCAGGGA 480

Query  481   TGGGATTAGTCTCCCCTGAGCTAGGAAGAATAGATGAAAATTCTATTAATACTTAATTAA 540
             ||||||||||||||||||||||||||||||||||||||||||||||||||||||||||||
Sbjct  481   TGGGATTAGTCTCCCCTGAGCTAGGAAGAATAGATGAAAATTCTATTAATACTTAATTAA 540

Query  541   CTAATTTT 548
             ||||||||
Sbjct  541   CTAATTTT 548


#-------------------------------------------------------------------
#-------------------------------------------------------------------
```

### 2.1.4 Multiple sequence alignments for the EhSINEs

Multiple sequence alignment (MSA) is used to align three or more sequences of similar length. There are so many tools for multiple sequence alignment but CLUSTALW is one of the best tools for aligning multiple nucleotide sequences. We have used CLUSTALW for aligning the sequences. When we align 23 cloned EhSINE1s, we found that the data can be classified into three categories: category I, a group of ten sequences which completely matched with a marked SINE; category II, a group of eight sequences which do not have tag and matching with genomic SINE copies; and category III, a group of five sequences which have 25-bp tag at the expected location but, matching with genomic SINEs rather than the marked SINE. In this way, we identified three groups. We have considered original SINE as a query and remaining clones (groups) as the databases. We have found mismatches using BLASTN. We are calling these groups as clusters and these clusters used as genotypes in quasispecies model.

### 2.1.5 Phylogenetic analysis of retrotransposons EhSINEs

Gblocks is used to eliminate poorly aligned positions and divergent regions of an alignment of DNA sequences. These positions need not be homologous. These positions may be saturated by multiple substitutions that’s why it is important to remove them prior to phylogenetic analysis. The important parameters of Gblocks are contiguous nonconserved positions, length of block and gap positions. The defaults values of these parameters are 8, 10 and none respectively. We are using the defaults values. Using Gblocks, we can reduce the necessity of manually editing multiple alignments. It makes the automation of phylogenetic analysis of large data sets workable and, finally, helps the reproduction of the alignments. One more important feature of this program is that, it is very fast in processing alignments. Therefore it is good for large-scale phylogenetic analyses. Trees were builded by SeaView (version 4.3.5). SeaView is a multiplatform, graphical user interface by which we can compute and draw parsimony, distance and PhyML phylogenetic trees. This can read and write various file formats of DNA sequences and phylogenetic trees, for example, FASTA, CLUSTAL, PHYLIP, Newick etc. We generate phylogenetic trees by parsimony method, using PHYLIP's dnapars algorithm and for this analysis we are using the parameters: randomize sequence order 5 times, ignore all gap sites, equally best trees retained 10000 and bootstrap with 100 replicates.

## 2.2 Quasipecies Model

Quasispecies are collection of genotypes that exist in a population at mutation selection balance. Quasispecies has been used in the context of large genome size but many concepts of quasispecies can be demonstrate in simple case of few genotypes which is our approach here. The model has been taken from (Bull, et al., 2005). This model is a concept of equilibia without initial condition of genotypes. Suppose there is a model (Bull, et al., 2005) which has two genotypes say $G_1$ and $G_2$ having fitness $f_1$ and $f_2$ where $f_1 > f_2$ and mutation rate $u_1$ and $u_2$ respectively (figure 7). Genotype $G_1$ has fitness $f_1$, and of those $f_1$ offspring a fraction $1-u_1$ retain the $G_1$ genotype. Its mutant will convert into other genotype $G_2$. All the mutants of genotype $G_2$ dies except its genotype which is reproduced with fidelity $1-u_2$. The product of fitness $f_1$ and the mutation-free fraction $(1 - u_1)$ of offspring i.e. $f_1(1 - u_1)$ is referred as the replacement rate of the genotype which is the number of $G_1$ offspring of a $G_1$ parent. We assume that this product number is greater than one (this assumption is not necessary but for our convenient we have assumed). Since $G_2$ has a lower fitness than $G_1$, it follows that, for small mutation rates $u_1$, the replacement rate of $G_2$ is lower than the replacement rate of $G_1$, i.e. $w_2(1-u_2) < w_1(1-u1)$. Therefore the replacement rate of $G_1$ exceeds that of $G_2$. $G_1$ will be maintained at equilibrium because of its higher replacement rate and $G_2$ because of mutation from $G_1$, so due to these two reasons both genotypes will maintained at stable point. If the mutation rate of $G_1$ increases such that $1\text{-}u_1$ of $G_1$ decreases relative to that of $G_2$, however, the replacement rate inequality will eventually reverse, giving a simple error catastrophe i.e. $G_1$ is no longer maintained. $G_1$ can not be recreated from $G_2$ because of lack of back mutations.

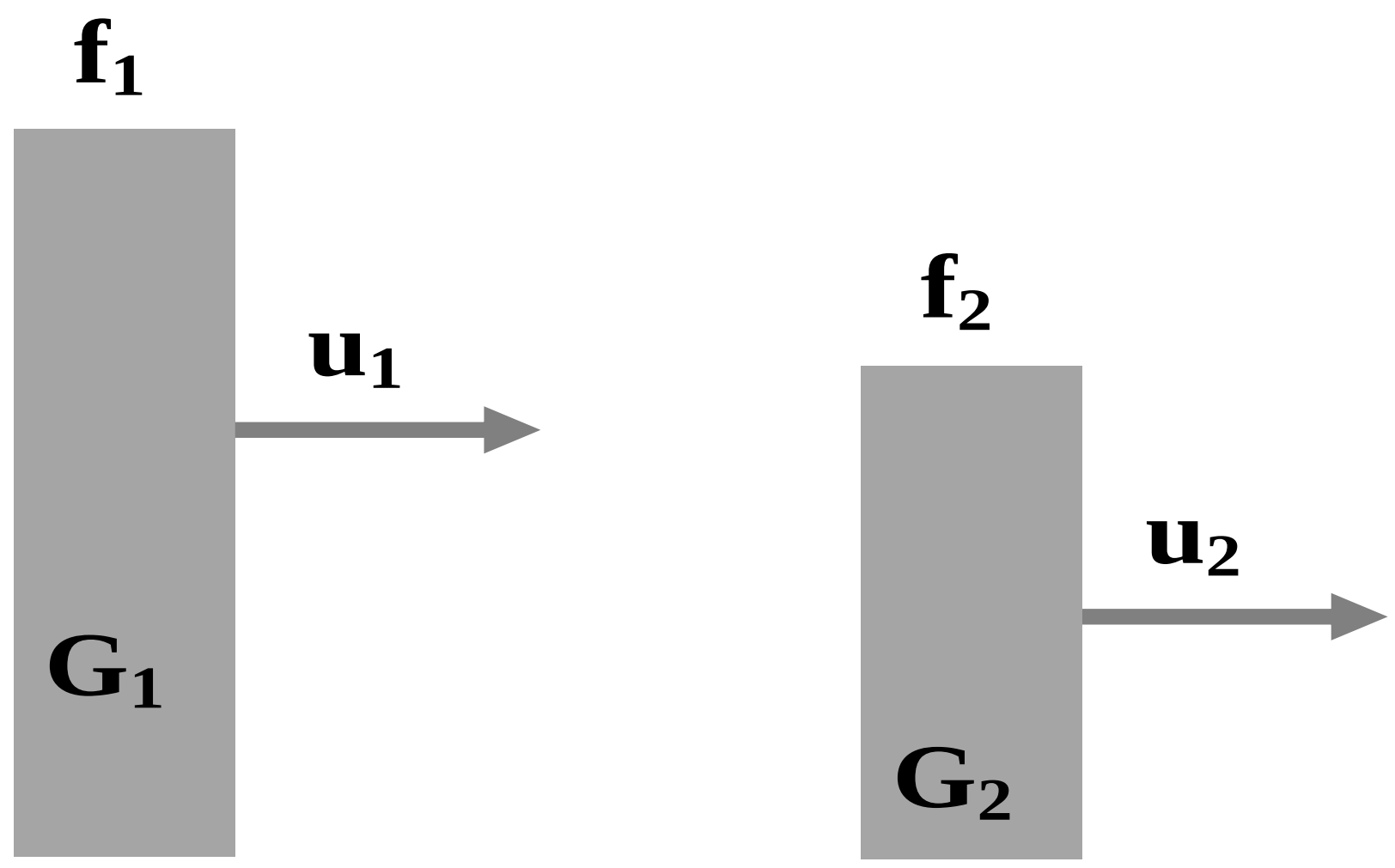

***Figure 7:*** *Model of two genotypes with forward mutation and fitness.*

*Each genotype $G_i$ has its own fitness $f_i$ and mutation rate $u_i$. Mutation is asymmetric, so that $G_1$ can recreate $G_2$, but not vice versa.*

Fitness landscape may contain multiple $G_i$ with fitness $f_i$ and mutation $u_i$ (Figure 8). Extinction take place at the point where the most adaptable genotype is unable to reproduce itself to maintain the a minimum population size ($f_i(1 - u_i) < 1$ for all genotypes).

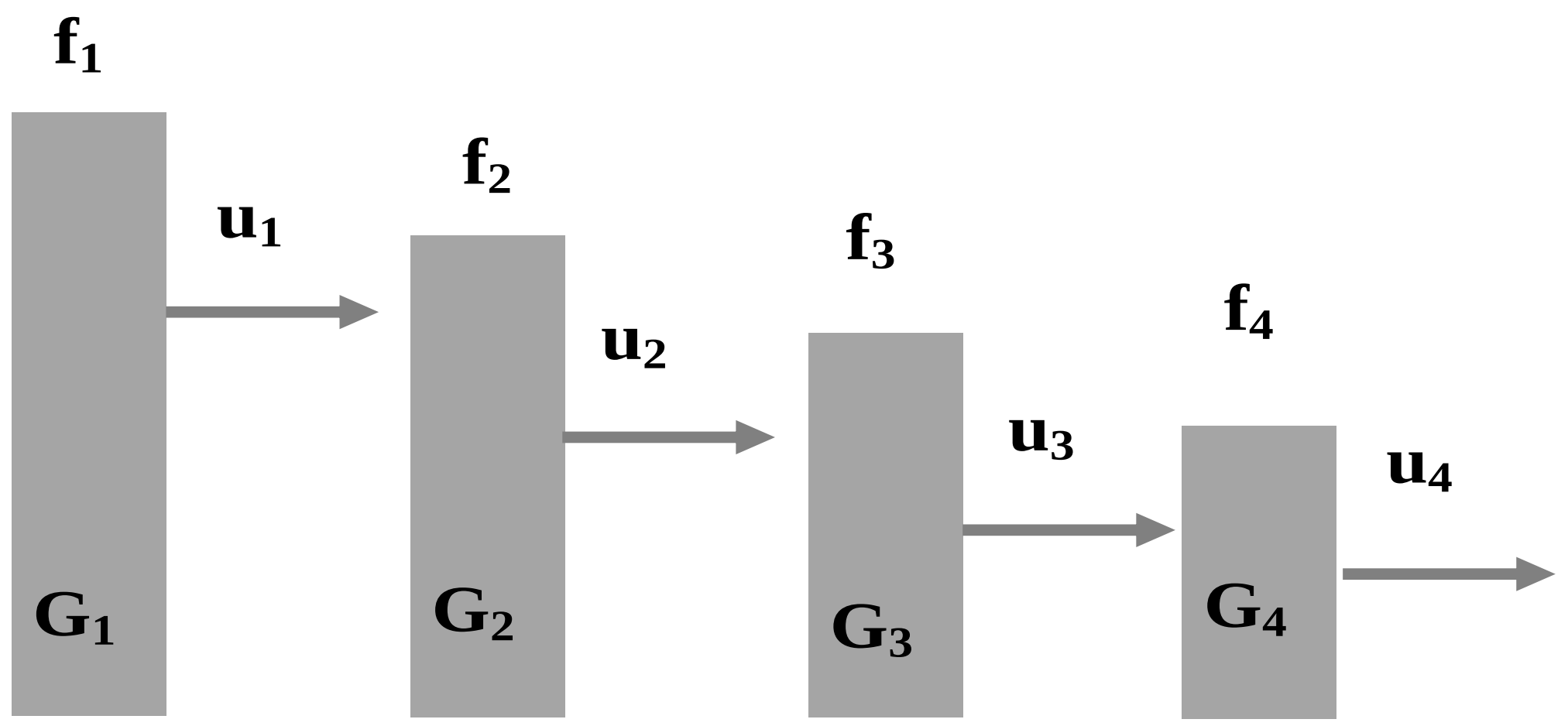

***Figure 8:*** *Model of several genotypes with forward mutation and fitness.*
Each genotype $G_i$ has its own fitness $f_i$ and mutation rate $u_i$.

There is no back mutation in the model of two genotypes, therefore $G_1$ will not regenerate once lost from the population (Figure 7). In the models of population genetics, back mutations are not often assumed (e.g. the infinite alleles and infinite sites models of population genetics (Kimura, et al., 1964; Ewens, et al., 1974). Since there are many systems where back mutation can occur therefore it is important to study its effects on the error threshold. The wild-type genome will not totally lose when back mutation occurs in system. Now consider a system where back mutation occurs (figure 9).

For the mathematical formulation consider a model (figure 9) which has back mutation $u_3$ as well as forward mutation $u_1$ and $u_2$. To explore the model, we use simple matrix algebra.

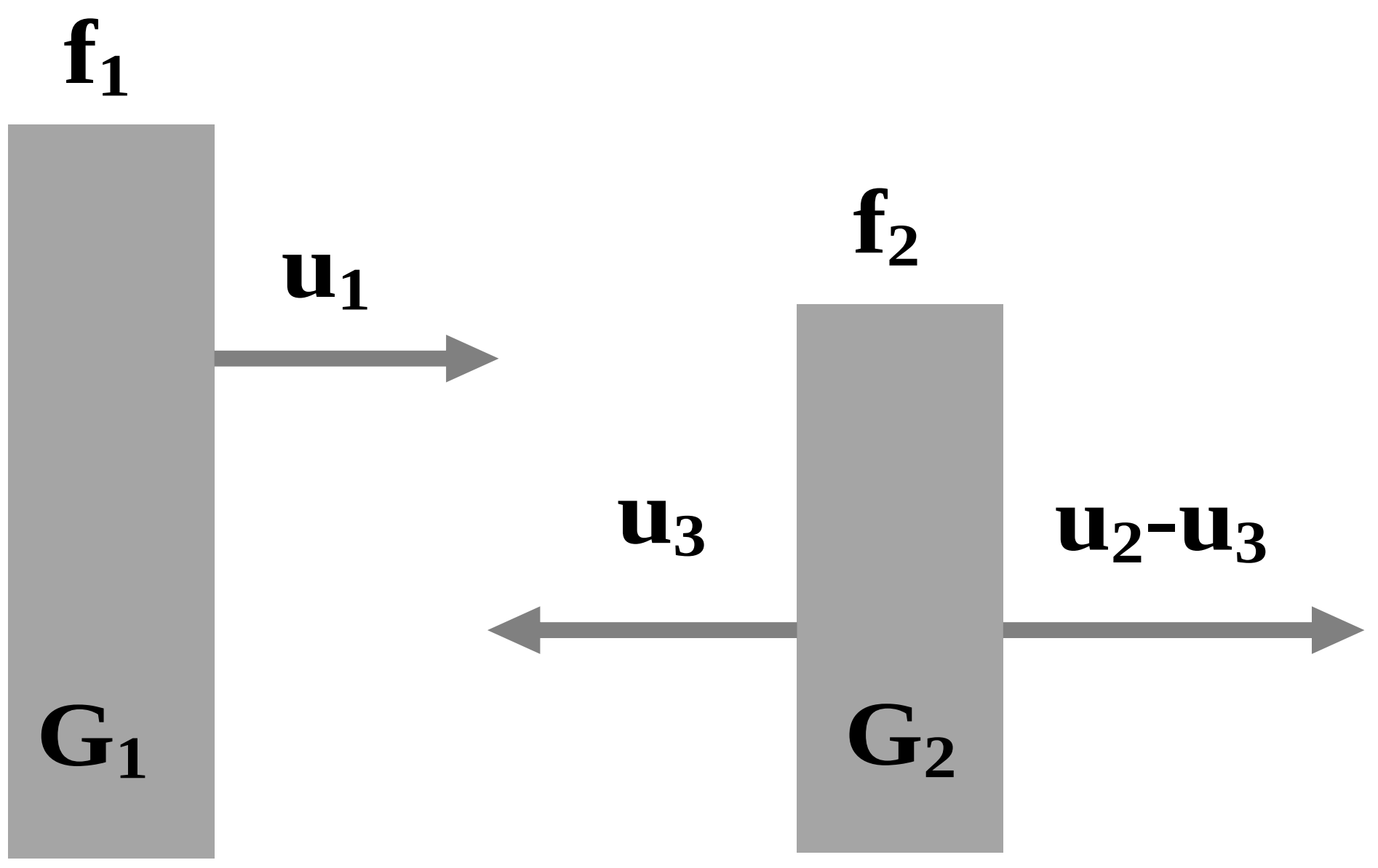

***Figure 9:*** *Model of two genotypes with back mutation, $u_3$.*
As in Figure, except that some of the mutations away from $G_2$ recreate genotype $G_1$. The total mutation rate of $G_2$ is still $u_2$, so the back mutation rate cannot exceed this value (i.e., $u_3 \leq u_2$).

.Consider discrete time steps, let $p_i$ be the number of individuals with genotype $G_i$. Two equations describe the system:

$$q_1' = q_1(f_1[1-u_1]) + q_2(f_1u_3) \qquad \text{...........(2.1)}$$

$$q_2' = q_1(f_2\, u_1) + q_2(f_2[1-u_2]) \qquad \text{...........(2.2)}$$

In the above equations,

$q_1$' represent the total individuals of genotype $G_1$ after one generation.

$q_2$' represent the total individuals of genotype $G_2$ after one generation.

First term of equation 1 i.e. $q_1(f_1[1-u_1])$ represents the individuals of $G_1$ which retain after mutation $u_1$ and second term tell us that the individuals comes from genotype $G_2$ with mutation rate $u_2$.

First term of equation 2 i.e. $q_1(f_2\, u_1)$ represents the individuals of $G_1$ which is converting into a genotype $G_2$ with mutation rate $u_1$ and second term i.e. $q_2(f_2[1-u_2])$ is a number of individuals which genotype $G_2$ retain after mutation.

The system of equation can be written in the form of transition matrix i.e.

$$\begin{pmatrix} q_1' \\ q_2' \end{pmatrix} = \begin{pmatrix} f_1[1-u_1] & f_1 u_3 \\ f_2 u_1 & f_2[1-u_2] \end{pmatrix} \begin{pmatrix} q_1 \\ q_2 \end{pmatrix}$$

Or $$(q_1', q_2')^T = M\,(q_1, q_2)^T$$

Where T denotes the transpose of a vector and

$$M = \begin{pmatrix} f_1[1-u_1] & f_1 u_3 \\ f_2 u_1 & f_2[1-u_2] \end{pmatrix} \qquad \text{..........(2.3)}$$

For the first generation we calculate $M(q_1, q_2)^T$ and for second generation calculation, initial condition will be the result of first generation i.e. $(q_1', q_2')$ and similarly we calculate for n generation. R and MATLAB softwares were used for programming.

If we take $u_3 = 0$, then equations (2.1) and (2.2) describe earlier model exactly (Figure 7). To find the behaviour of system for the long time, we assume $q_1(t)$ and $q_2(t)$ as t (generation time) becomes arbitrary large. Finding eigen values is a standard method to know the behaviour of such linear system. The characteristic equation of M can be given by |M – kI| and k is called eigen values if determinant of (M – kI) equals to zero i.e.

$$|M - kI| = 0$$

This implies,

$$k^2 - k(f_1[1 - u_1] + f_2[1 - u_2]) + ( f_1f_2[1 - u1][1 - u2] - f_1f_2u_1u_2) = 0 \qquad \text{.........(2.4)}$$

If there is no back mutation then it gives previous model (Figure 7) and in this case M will be a lower triangular matrix and the eigen values of this matrix are

$$k_1 = f_1[1\text{-}u_1] \text{ and } k_2 = f_2[1\text{-}u_2]$$

These two eigen values $k_1$ and $k_2$ are the replacement rate of genotype $G_1$ and $G_2$ respectively.

If back mutation $u_3 > 0$, then the eigen values are

$$k_1 = \frac{a + b + \sqrt{(a-b)^2 + 4f_1f_2u_1u_3}}{2}$$

$$k_2 = \frac{a + b - \sqrt{(a-b)^2 + 4f_1f_2u_1u_3}}{2}$$

Where, $a = w_1(1 - u_1)$ and $b = w_2(1 - u_2)$.

If there is no back mutation then a transition from one eigen value ($k_1$) to another ($k_2$) as the dominant eigenvalue found at a = b, therefore for any value of a and b, $k_1$ remains superior. As $u_3 \rightarrow 0$, the eigenvalues converge to each other but do not cross. Hence, error threshold does not exist until and unless $u_3$ is not equal to zero. Yet system appears more and more similar in both cases: $u_3$ approaching to zero and equals to zero. So the mathematical discontinuity between systems which have or do not have back mutation, is not easily translate into a meaningful biological differentiation, but excess of back mutation may change whole scenario.

### 2.2.1 Model parameters

As we have discussed above, our model have two parameters, mutation rate and fitness. Mutation rate is a very important parameter for understanding the genetic structure and evaluation of population over time (Drake, et al., 1999).

### Method for finding mutation rate of clustered sequences

To find the mutation rate of retrotransposable elements of *E. histolytica,* we have used the methods of previous published work (Drake, et al., 1999). When mutation rates are calculated per genome per genome replication, different broad groups of organisms shows different values (Drake, et al., 1998), roughly 0.2 for retroelements (Drake, et al., 1999). A simple scheme of cell infection is shown in figure 10. One or more virus particles can infect a cell. Each infecting sequence is copied iteratively to produce complimentary strands which in turn get copied iteratively. In this way, final strands are same as the infecting strands. We assume that, within a single infection cycle, the final strands will never or rarely re-enter the starting of the cycle.

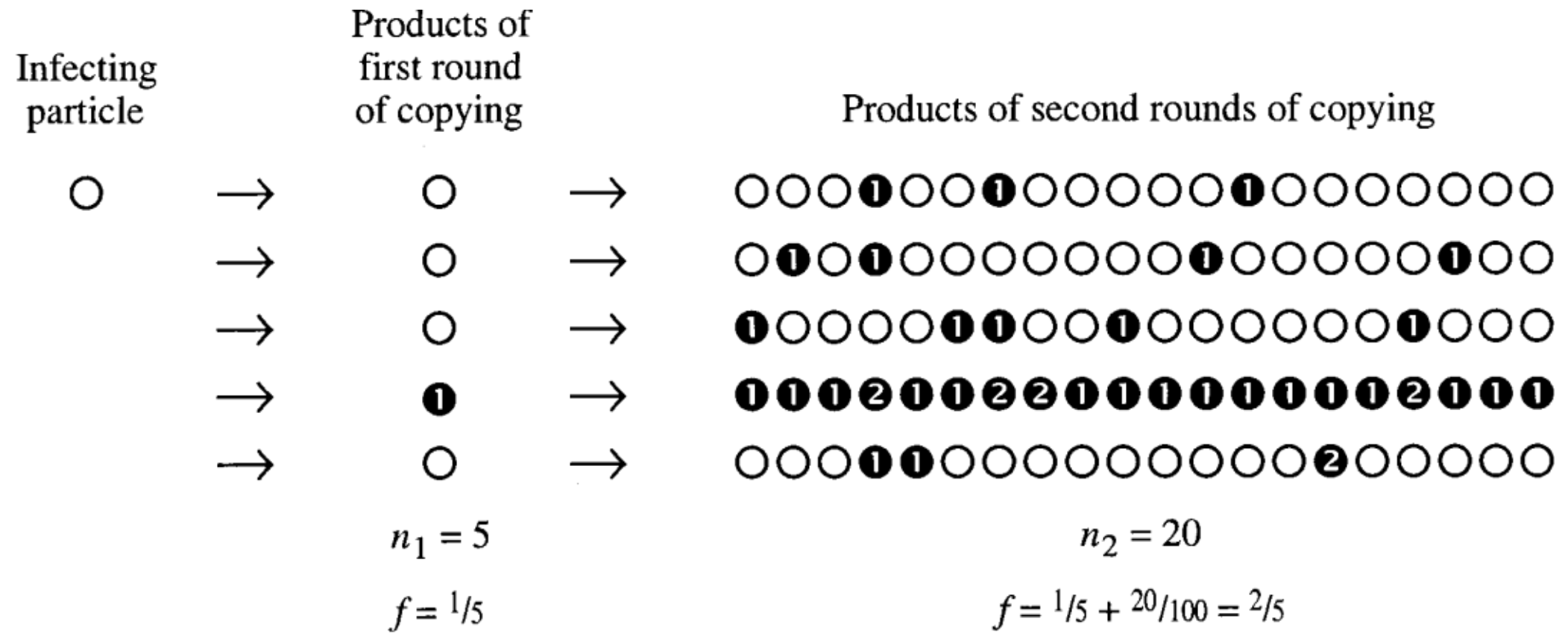

***Figure 10:*** *The accumulation of mutations during a single round of infection.*

The diagram shows the consequences of an arbitrary $u_g = 0.2$ at both the first and second rounds, and the scheme is simplified by setting $n_2 = 20$ for all second rounds of copying. The burst size is $n_1 n_2 = 100$. ○, unmutated sequence; ● , sequence with mutation. (Adapted from Drake et al., 1999)

For a given mutation frequency, mutation rate can be easily calculated. Mutation rate per base can be calculated using the method (Drake, et al., 1999) as

$u_b$ = 1.462×[(all mutations)/(base pair substitutions)] × [the number of bases at which the event can occur].

i.e. $$u_b = 1.462f/2cN \quad \ldots\ldots\ldots..(2.6)$$

Mutation rate per genome per replication $u_g$ can be determined by multiplying genome size in equation (2.6). Therefore,

$$u_g = 1.462fG/2cN \quad \ldots\ldots\ldots..(2.7)$$

In the above equation (2.7),

f is mutation frequency.

G is genome size.

c is growth parameter of retroviruses.

N is mutational target size (i.e. A→T or C or G) therefore N= 1/3.

Example of mutation rate calculation,

Suppose a virus have a genome of length 7,124 bases and mutation frequency is 1/6457 = $1.55 \times 10^{-4}$. Mutational target site N = 1/3 and c = 2.5 then,

$$u_g = (1.462 \times 1.55 \times 10^{-4} \times 7{,}124)/(2 \times 2.5 \times 1/3)$$

This implies $u_g = 0.968$

## Fitness

Fitness describes the ability to both survive and reproduce. We have taken fitness from published work and also assigned different values for calculation (Bull et al., 2005) for model.

# Chapter 3

# Results and Discussions

In this chapter we present results from both datasets of retrotransposons (SINE1s) of *E. histolytica*. We have calculated the mutation rate of EhSINE1s for both datasets and drawn phylogenetic tree for newly determined EhSINE1s (dataset II). We have also discussed the variation in lengths of EhSINE1s for both datasets.

Using the quasispecies model, we have shown how sequences of SINE1s vary within the population. The outputs of quasispecies model is discussed in the presence and the absence of back mutation by taking different values of fitness.

## 3.1 Analysis of data taken from previous published work

There are many families of repetitive elements which are present in the genome of *E. histolytica.* Among these the retrotransposons (EhSINE1s) were examined in detail. There are many aspects of the biology of the parasite *E. histolytica* which are not clear, specially why it is pathogenic while *E. dispar* is not, the great genetic diversity determined amongst patient isolates. Mobile genetic elements, having ability to alter gene expression, may be important to solve these puzzles. In this section, we will discuss about the results of dataset I.

### 3.1.1 Length variation in EhSINE1s

Three hundred and ninety three sequences of EhSINE1 were extracted. This number is similar to that reported in a previous work (Lorenzi, et al., 2010; Huntley, et al., 2010). Out of these 393 EhSINE1s, 158 EhSINE1s had one repeat, 67 two repeats, 7 three repeats and 3 four repeats (Huntley, et al., 2010), 95 EhSINE1s of the appropriate length for 3-repeat ones and 63 EhSINE1s had no recognisable repeats at all (Huntley, et al., 2010).The distribution of lengths of the 393 EhSINE1s are shown in figure 11. It can be easily seen that the range of the lengths of these 393 SINE1s are from 451 to 684 bp but they are not normally distributed. There are 26 sequences which are 548 bp in length, the longest observed among all EhSINE1s.

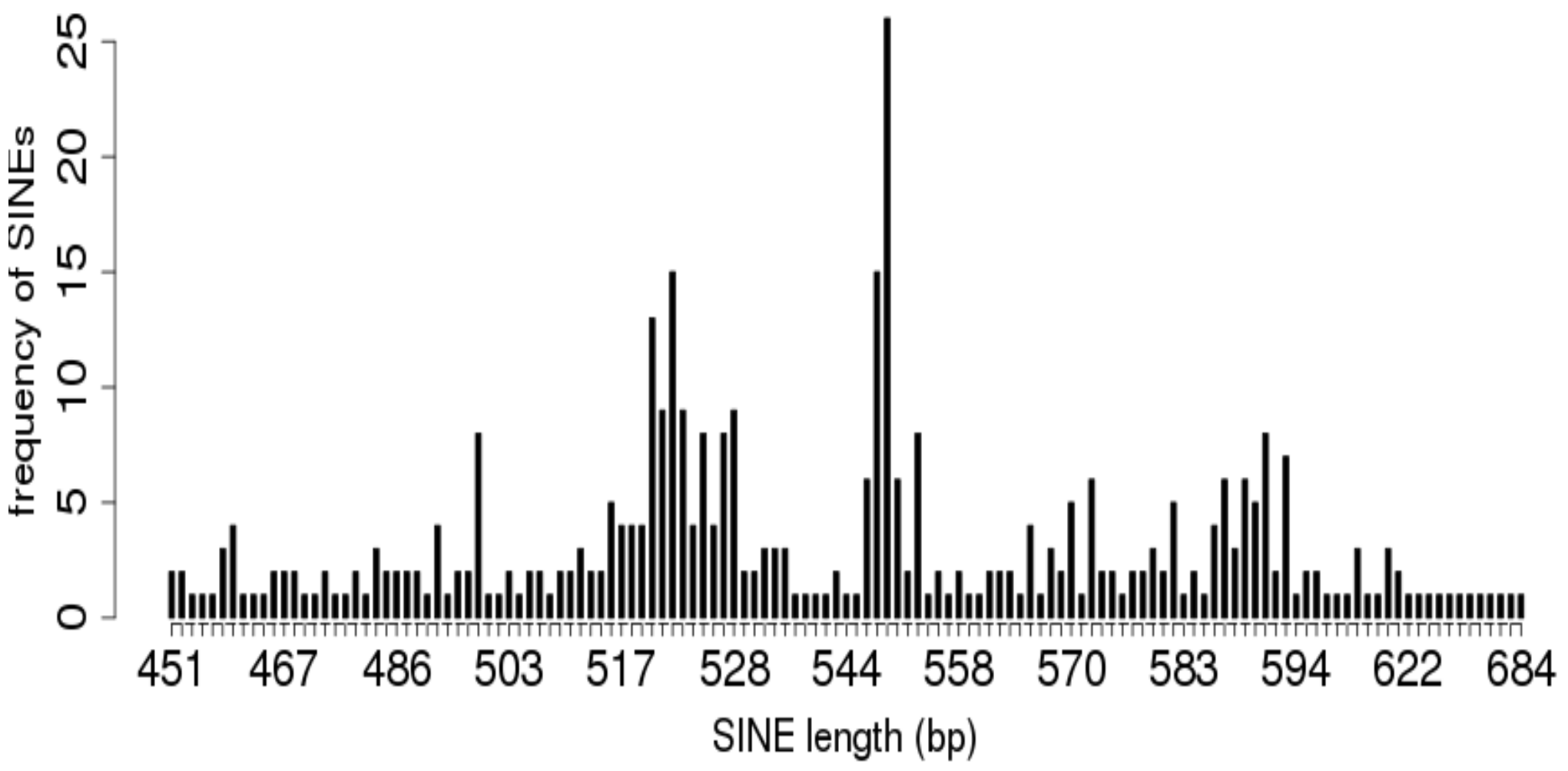

***Figure 11***: *Distribution of Lengths of 393 EhSINE1s*

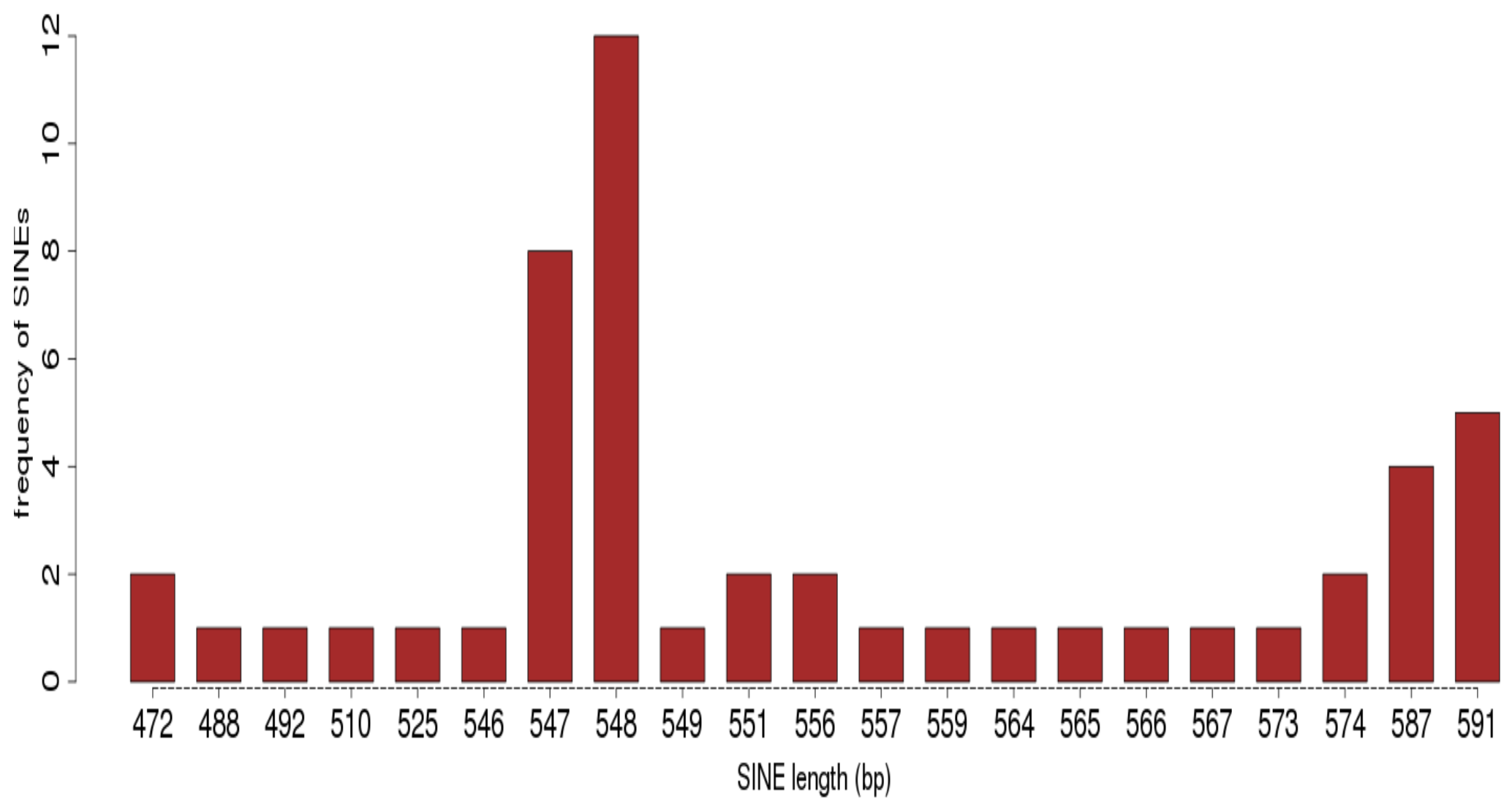

***Figure 12:*** *Distribution of Lengths of 50 EhSINE1s*

We identified 50 EhSINE1s, divided into 8 sets of "identical" (≥ 98 % sequence identity and 80% query coverage) EhSINE1s - one set each containing 18, 6, 6, 6, 5, 4, 4, 4 and 3 members. Figure 12 shows the distribution of length of 50 sequences and these were taken for the analysis.

### 3.1.2 Mutation rate of EhSINE1s

Mutation rate of sequences can be calculated easily. Let's take an example in genotype $G_1$, query sequence (gi|DS572593|SP-EP:697-150|548) is 548 bp long, lets call it genome size G and subject sequence (gi|DS571423|SP-EP:1030-483|548) is 548 bp long and there are 4 mismatches. Therefore the mutation frequency is 4/548(subject sequence length). We have considered the growth parameter c = 7.5 (Drake, et al., 1999) and mutational target site N = 1/3. Mutation rate per base is given by

$$u_b = 1.462f/2cN$$

Therefore, $u_b = (1.462\times 4/548) / (2\times 7.5\times 1/3)$

$$u_b = 17.54/ 8220$$

$$u_b = 0.0021$$

Calculation of mutation rate per genome per replication is given by

$$u_g = 1.462fG/2cN$$

Therefore, $u_g = 0.00223\times 548$

$$u_g = 1.17$$

In this way, mutation rates of all subject sequences against query sequence calculated in genotype $G_1$. This whole process is repeated to find mutation rates of all sequences in all genotypes ($G_1$, $G_2$, $G_3$, $G_4$, $G_5$, $G_6$, $G_7$, and $G_8$) and results are shown in Table 4.

**Table 4.** Mutation rates per base pair ($u_b$) and mutation rates per genome per replication ($u_g$) of 50 sequences.

| G | Q. Id | S. Id | % Id | A.L. | m | $u_b$ | $u_g$ |
|---|---|---|---|---|---|---|---|
| $G_1$ | gi\|DS572593\|SP-EP:697-150\|548 | gi\|DS571423\|SP-EP:1030-483\|548 | 99.27 | 548 | 4 | 0.0021 | 1.17 |
| | | gi\|DS571272\|SP-EP:1423-1970\|548 | 98.91 | 548 | 6 | 0.0032 | 1.75 |
| | | gi\|DS571382\|SP-EP:18914-18367\|548 | 98.72 | 548 | 7 | 0.0037 | 2.04 |
| | | gi\|DS571151\|SP- | 98.73 | 550 | 3 | 0.0016 | 0.88 |
| | | EP:154899-154352\|548 | | | | | |
| | | gi\|DS571317\|SP-EP:27553-28099\|547 | 98.54 | 549 | 5 | 0.0027 | 1.46 |
| | | gi\|DS571347\|SP-EP:7763-8309\|547 | 98.54 | 548 | 7 | 0.0037 | 2.05 |
| | | gi\|DS571347\|SP-EP:2150-2696\|547 | 98.54 | 548 | 7 | 0.0037 | 2.05 |
| | | gi\|DS571247\|SP-EP:15177-14629\|549 | 98.36 | 549 | 8 | 0.0043 | 2.33 |
| | | gi\|DS571278\|SP-EP:6338-5791\|548 | 98.18 | 549 | 8 | 0.0043 | 2.34 |
| | | gi\|DS571145\|SP-EP:96823-97370\|548 | 98.18 | 548 | 10 | 0.0053 | 2.92 |
| | | gi\|DS571179\|SP-EP:37873-38420\|548 | 98.18 | 548 | 10 | 0.0053 | 2.92 |
| | | gi\|DS571158\|SP-EP:27154-27704\|551 | 98.01 | 552 | 6 | 0.0032 | 1.74 |
| | | gi\|DS571257\|SP-EP:45861-46408\|548 | 98.00 | 550 | 7 | 0.0037 | 2.05 |
| | | gi\|DS571508\|SP-EP:4500-3954\|547 | 98.00 | 549 | 8 | 0.0043 | 2.34 |
| | | gi\|DS571494\|SP-EP:2811-3357\|547 | 98.00 | 549 | 8 | 0.0043 | 2.34 |
| | | gi\|DS571190\|SP-EP:67757-68303\|547 | 98.00 | 550 | 6 | 0.0032 | 1.76 |
| | | gi\|DS571346\|SP-EP:26099-25553\|547 | 98.10 | 527 | 10 | 0.0053 | 2.93 |
| $G_2$ | gi\|DS571207\|SP-EP:6989-6399\|591 | gi\|DS571380\|SP-EP:2811-2221\|591 | 100.00 | 591 | 0 | 0 | 0 |
| | | gi\|DS571661\|SP-EP:3247-2657\|591 | 100.00 | 591 | 0 | 0 | 0 |
| | | gi\|DS571159\|SP-EP:21722-21132\|591 | 99.83 | 591 | 1 | 0.0005 | 0.29 |
| | | gi\|DS571481\|SP-EP:868-278\|591 | 99.66 | 591 | 2 | 0.0010 | 0.58 |
| | | gi\|DS571828\|SP-EP:3562-3038\|525 | 99.38 | 483 | 3 | 0.0017 | 0.99 |
| $G_3$ | gi\|DS571310\|SP-EP:11751-11204\|548 | gi\|DS571255\|SP-EP:5733-5183\|551 | 98.55 | 551 | 5 | 0.0027 | 1.45 |
| | | gi\|DS571147\|SP-EP:112278-111732\|547 | 98.54 | 548 | 7 | 0.0037 | 2.05 |
| | | gi\|DS572093\|SP-EP:1734-1187\|548 | 98.36 | 548 | 9 | 0.0048 | 2.63 |
| | | gi\|DS571487\|SP-EP:12336-12883\|548 | 98.36 | 548 | 9 | 0.0048 | 2.63 |
| | | gi\|DS571277\|SP-EP:39116-38571\|546 | 98.18 | 548 | 8 | 0.0043 | 2.35 |
| $G_4$ | gi\|DS571375\|SP-EP:11554-12140\|587 | gi\|DS571601\|SP-EP:8125-7539\|587 | 98.64 | 588 | 6 | 0.0030 | 1.75 |
| | | gi\|DS571499\|SP-EP:10454-9868\|587 | 98.64 | 588 | 6 | 0.0031 | 1.75 |
| | | gi\|DS571333\|SP-EP:24586-25172\|587 | 98.13 | 587 | 11 | 0.0055 | 3.21 |
| | | gi\|DS572494\|SP-EP:1-559\|559 | 98.75 | 560 | 5 | 0.0026 | 1.54 |
| $G_5$ | gi\|DS571897\|SP-EP:627-1193\|567 | gi\|DS571208\|SP-EP:64860-64295\|566 | 98.24 | 568 | 7 | 0.0036 | 2.05 |
| | | gi\|DS571442\|SP-EP:14666-15230\|565 | 98.24 | 567 | 8 | 0.0041 | 2.35 |
| | | gi\|DS571668\|SP-EP:993-1556\|564 | 98.24 | 567 | 7 | 0.0036 | 2.06 |
| $G_6$ | gi\|DS571874\|SP-EP:2490-3045\|556 | gi\|DS571827\|SP-EP:3250-2695\|556 | 99.82 | 556 | 1 | 0.0005 | 0.29 |
| | | gi\|DS572564\|SP-EP:309-865\|557 | 98.03 | 557 | 10 | 0.0052 | 2.92 |
| | | gi\|DS572142\|SP-EP:1214-1723\|510 | 99.61 | 507 | 1 | 0.0006 | 0.32 |
| $G_7$ | gi\|DS572182\|SP-EP:493-964\|472 | gi\|DS571675\|SP-EP:2982-2491\|492 | 99.58 | 472 | 2 | 0.0012 | 0.56 |
| | | gi\|DS571660\|SP-EP:757-270\|488 | 99.79 | 469 | 0 | 0 | 0 |
| | | gi\|DS572664\|SP-EP:830-359\|472 | 99.36 | 472 | 3 | 0.0019 | 0.88 |
| $G_8$ | gi\|DS571250\|SP-EP:22245-21673\|573 | gi\|DS571250\|SP-EP:42224-41651\|574 | 99.13 | 572 | 4 | 0.0020 | 1.18 |
| | | gi\|DS571151\|SP-EP:120325-119752\|574 | 98.08 | 574 | 10 | 0.0051 | 2.92 |

Abbreviations (genotypes (G), query sequences (Q.Id), subject sequences (S.Id), sequence similarities (% Id), alignment lengths (A.L.), Mismatches (m)).

**Calculation of mutation rate of genotypes:**

Let's M is the mean of mutation rates of all subject sequences against query sequence in genotype $G_1$.

$$M = 35.11/17$$

$$M = 2.06$$

Similarly mean of mutation rates of all subject sequences against query sequences in all genotypes ($G_2$, $G_3$, $G_4$, $G_5$, $G_6$, $G_7$ and $G_8$) have been calculated and the corresponding values were 0.372, 2.22, 2.06, 2.15, 1.17, 0.47 and 2.05 respectively.

Now the probability of conversion of offsprings from one genotype to other genotype is given by

$$u_i = \frac{\text{Mutation rate of particular genotype}}{\text{Total sum of mutation rate of all genotypes}}$$

For example, mutation rate of $G_1$ is given by

$$u_1 = 2.06/12.55$$

$$u_1 = 0.16$$

Similarly the mutation rates of all genotypes were calculated and shown in Table 5.

**Table 5.** Showing genotypes ($G_1$, $G_2$,.. $G_8$) and their corresponding mutation rate $u_i$.

| Genotype | Mutation rate $u_i$ |
|---|---|
| $G_1$ | 0.16 |
| $G_2$ | 0.03 |
| $G_3$ | 0.17 |
| $G_4$ | 0.16 |
| $G_5$ | 0.17 |
| $G_6$ | 0.09 |
| $G_7$ | 0.04 |
| $G_8$ | 0.16 |

### 3.1.3 Sequence variation within the population of EhSINE1s

To see to the variation in EhSINE1s, we have consider four genotypes $G_1$, $G_2$, $G_3$ and $G_6$ having mutation rates 0.16, 0.03, 0.17 and 0.09 (Table 5) and fitness 0.9, 0.7, 0.5, 0.2 (chosen randomly in decreasing order) respectively. Results of sequence variation using quasispecies model are shown in figure 13. X-axis shows the generation of population and y-axis shows the frequency of population of the genotypes. It can be easily seen from figure 13 that the population of all genotypes exists for few generations but $G_{6,}$ having lowest replacement rate (i.e. $f_i(1-u_i)$ = 0.182, other values 0.756, 0.679 and 0.415 for $G_1$, $G_2$ and $G_3$ respectively) would not stay in populations for further generation. The genotype $G_1$ and $G_2$ holds approximately 72 % and 28% population respectively (figure 13) and these are growing constantly after 20 generations. Therefore it is easy to say that the population of all genotypes may exist but frequency will differ and it is a property of quasispecies theory.

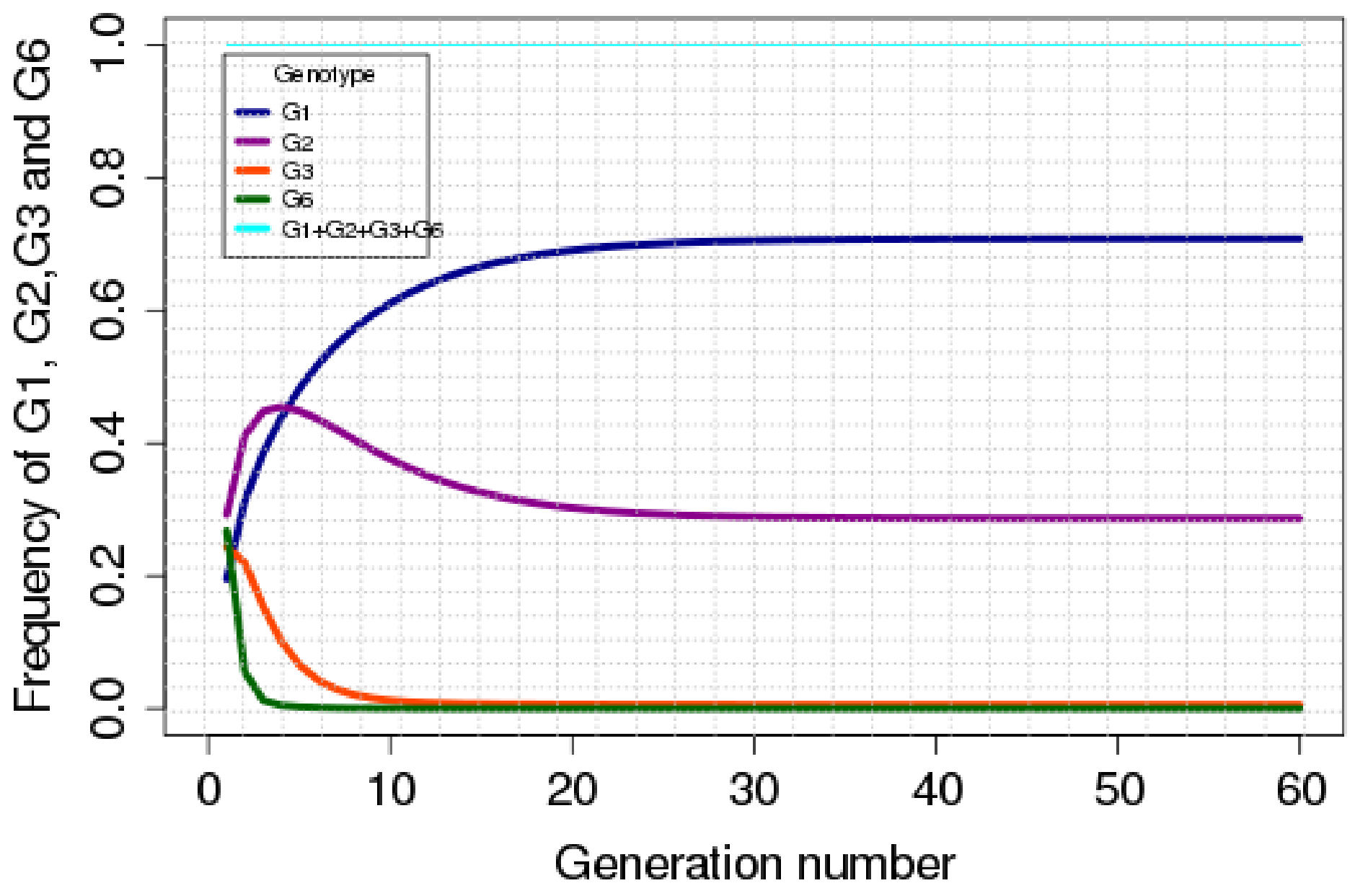

***Figure 13:*** *Variation of EhSINE1s with time generation*

### 3.1.4 Results using Quasispecies model without back mutation

Quasispecies model is independent of its initial conditions i.e. whatever initial conditions we take, the final distribution of population of genotypes, meet at equilibrium point (stable state). Consider a quasispecies model for two genotypes in which one genotype $G_1$ (say) (population size $p_1$ = 3000) fully dominant to the other genotype $G_2$ (population size $p_2$ = 20). Consider the same model again but now $G_2$ dominant ($p_2$=2000) to $G_1$ ($p_1$ = 15). Using the above population size; we describe the results of EhSINE1s using quasispecies model for different values of fitness.

**Result for fitness 1.5 and 1:**

Consider a model with forward mutation rates $u_1$=0.16 and $u_2$=0.03 (from table 5), and fitnesses $f_1$ = 1.5 and $f_2$ = 1 (Bull, et al., 2005). The figures (14, 15) are drawn using these parameters. For genotype $G_1$, the upper, dark green curve represents a different starting condition other than the lower dark magenta curve, but both equilibrate to the same value (around 6.3) after 35 generations (figure 14). For genotype $G_2$, both curves equilibrate to the same frequency value but this value differs from the equilibration frequency of $G_1$ (figure 15).

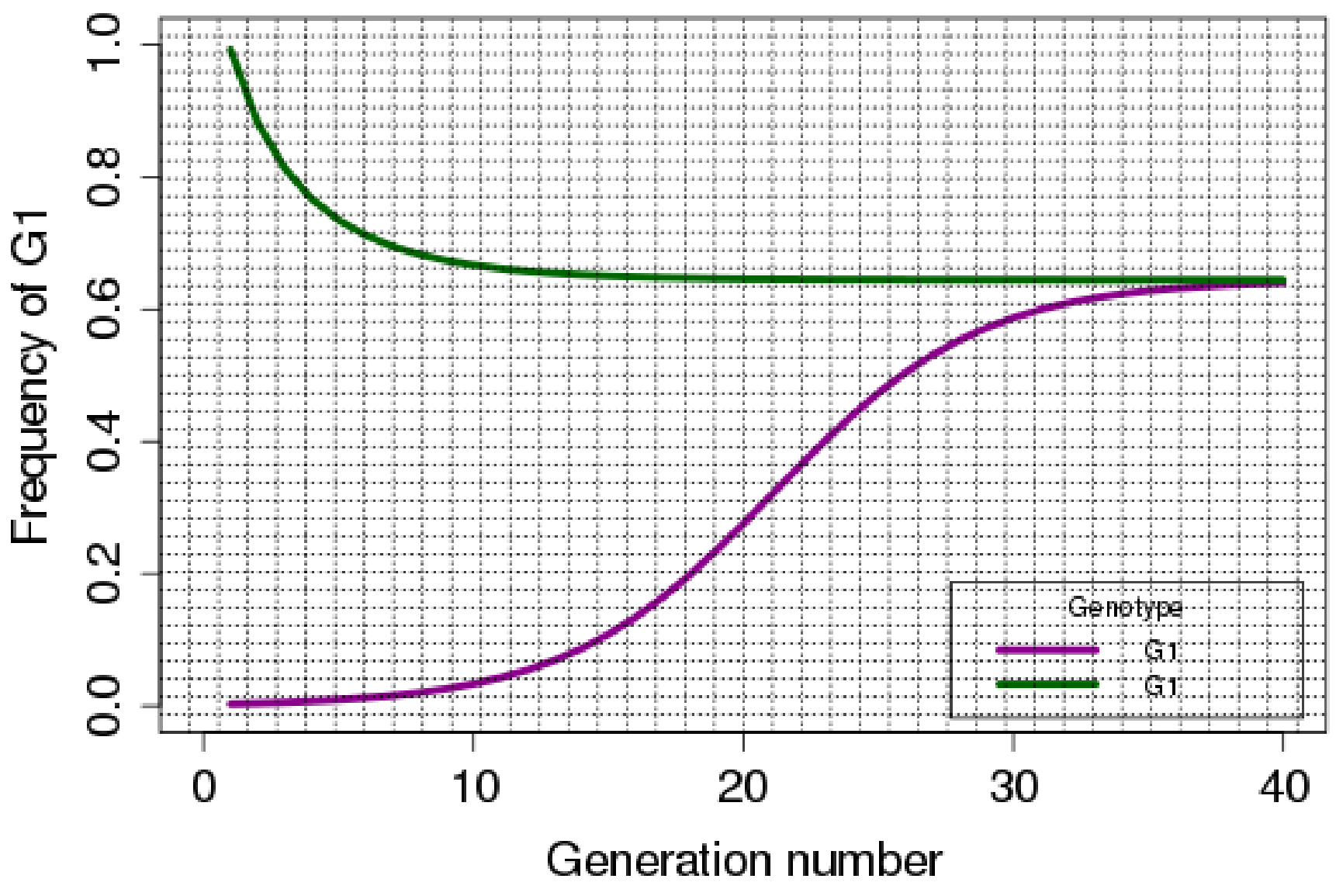

***Figure 14:*** *Dynamic approach to the quasispecies equilibrium, for genotype $G_1$ with different initial condition.*

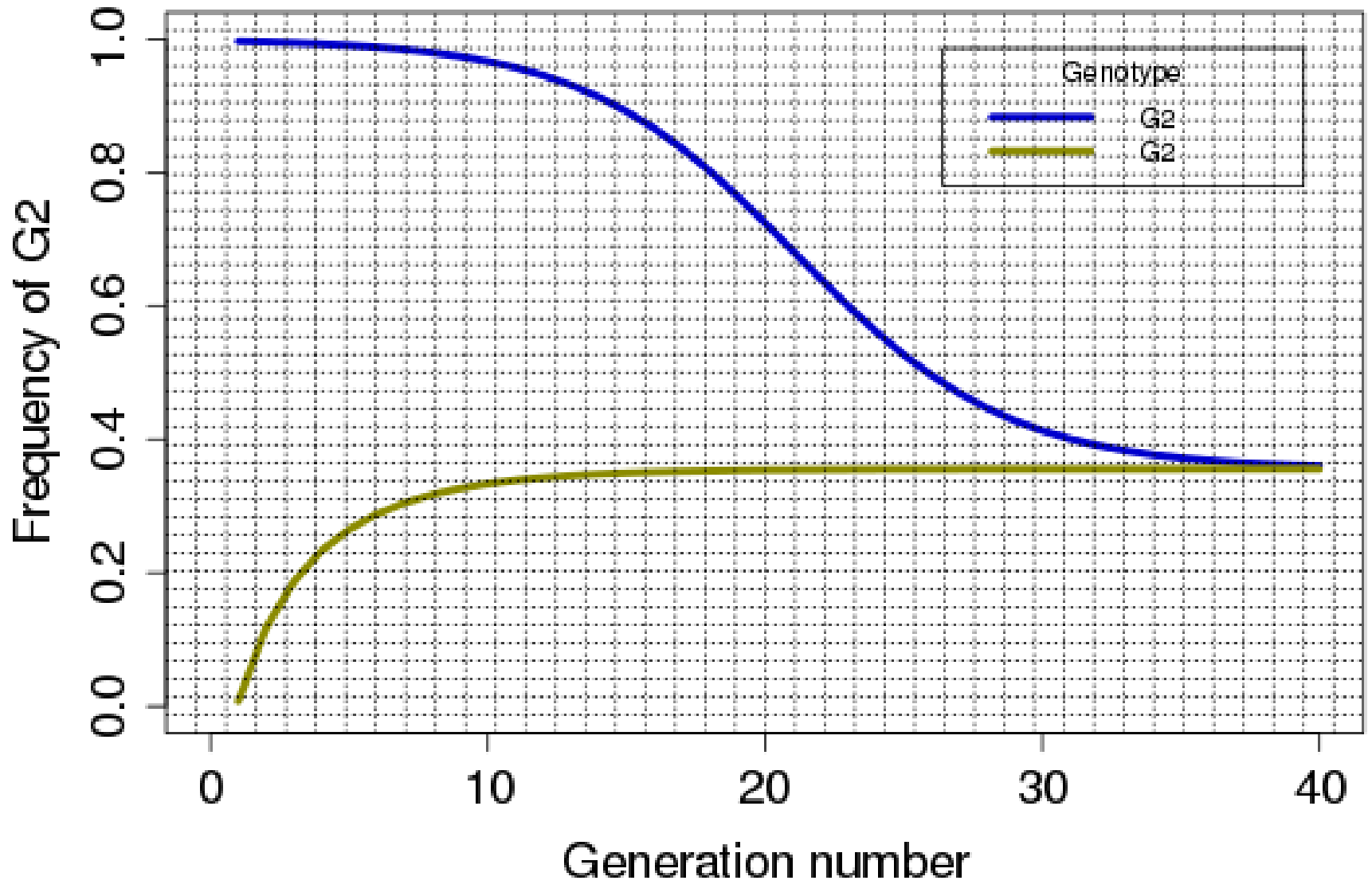

***Figure 15:*** *Dynamic Approach to the quasispecies equilibrium, for genotype $G_2$ with different initial conditions.*

**Results for fitnesses 0.5 and 1**:

Consider a model with two genotypes $G_1$ and $G_3$ with corresponding mutation rate $u_1$=0.16 and $u_3$ = 0.17 ($u_1 < u_3$), and fitness $f_1$ = 0.5 and $f_2$ = 1.0 ($f_1 > f_2$) (randomly chosen). Figure is drawn using quasispecies model with above parameters. It can be seen from the figure that one genotype ($G_1$) extinct after 7 generations (figure 16) i.e. $G_1$ is no longer active whereas only genotype $G_3$ exist (figure 17) and therefore these results indicate the definition of error catastrophe.

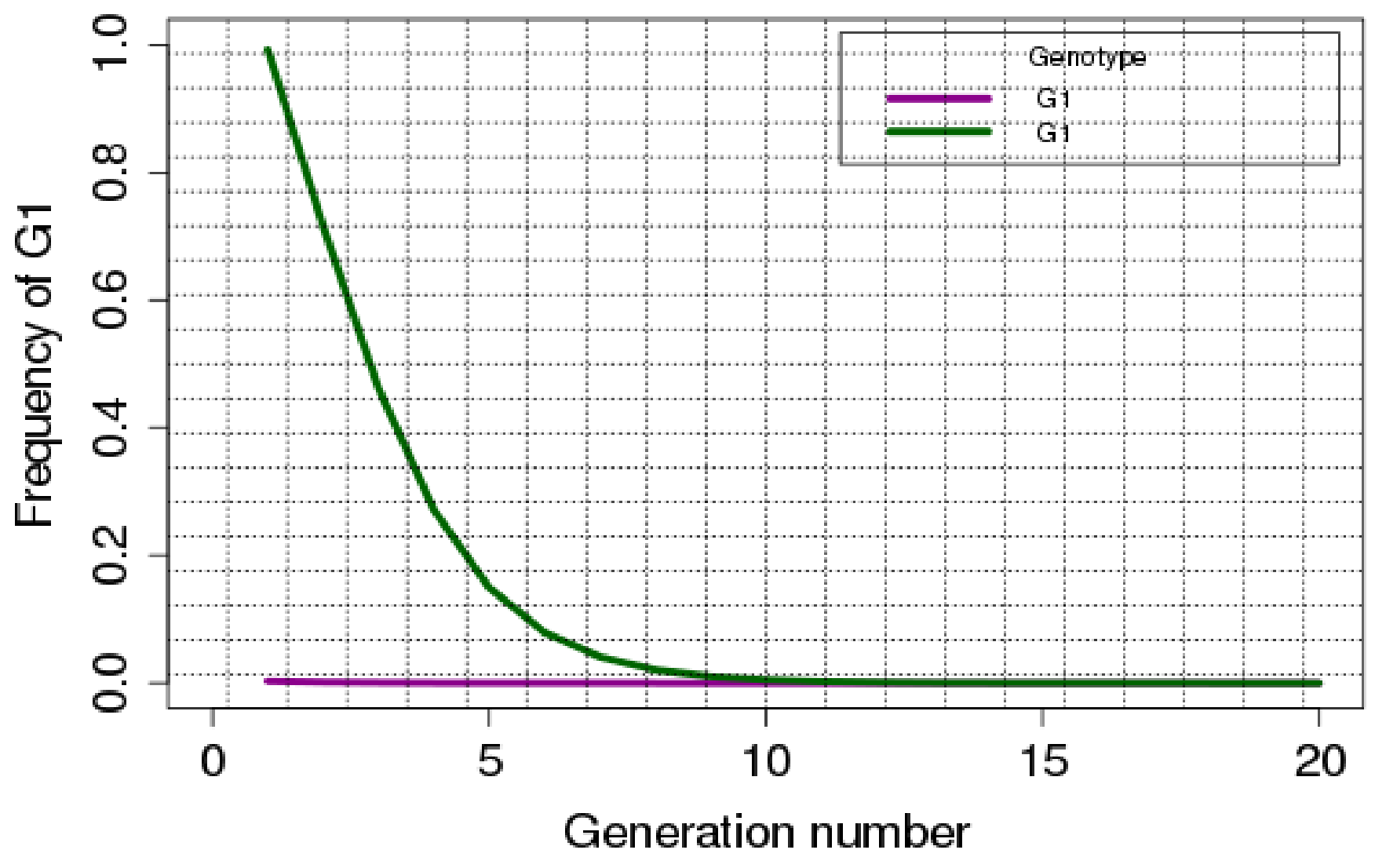

***Figure 16:*** *Showing error catastrophe, Extinction of genotype $G_1$*

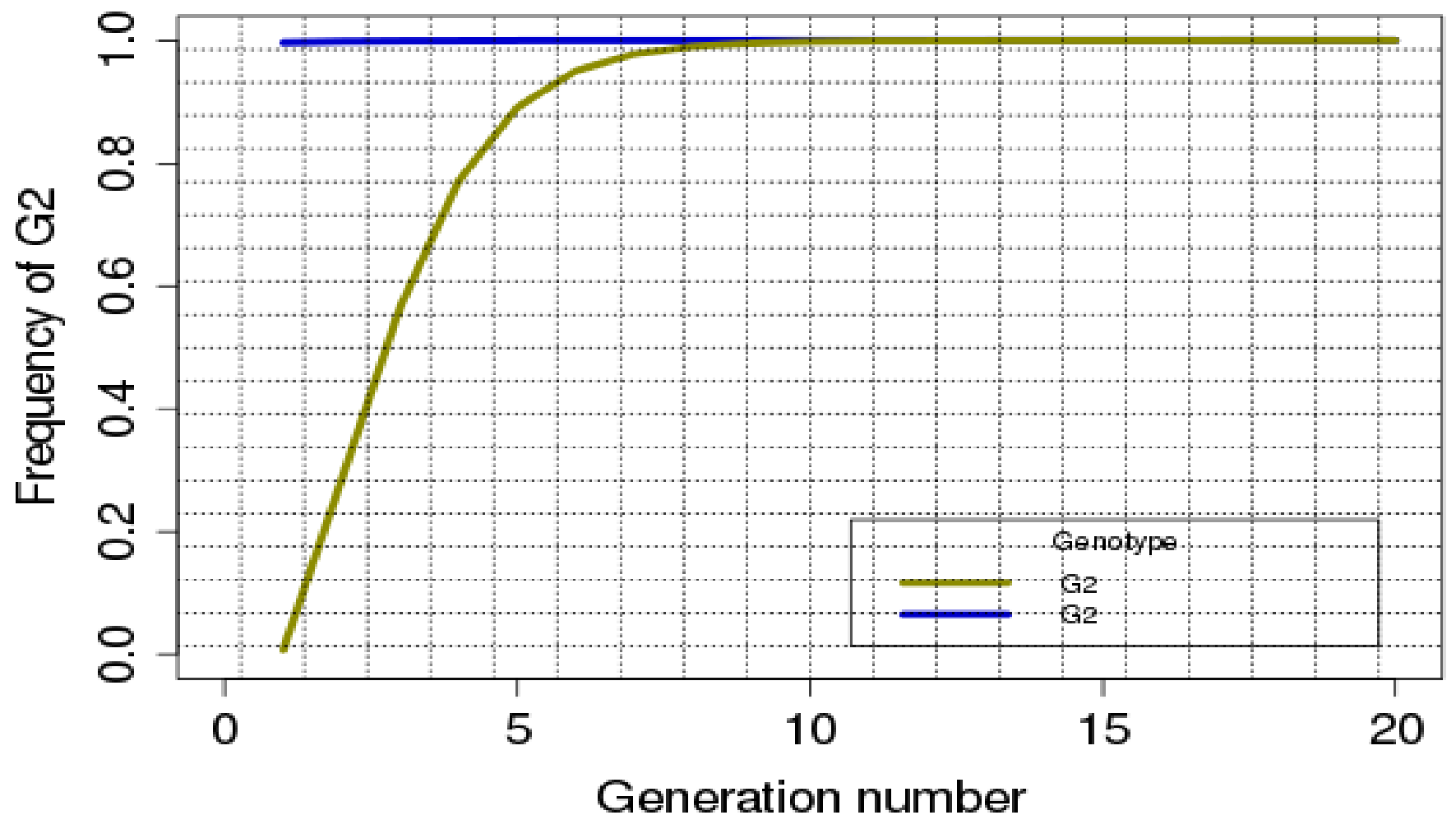

***Figure 17:*** *Showing error catastrophe where only genotype $G_3$ present*

**Result for fitness 0.7 and 0.4:**

Consider a model with two genotypes $G_1$ and $G_2$ having mutation rate $u_1$=0.16 and $u_2$ = 0.03 ($u_1 > u_2$), and fitness $f_1$= 0.7 and $f_2$ = 0.4 ($f_1 > f_2$) (randomly chosen). We can observe directly from figures (18, 19) that $G_1$ equilibrate at same value (around 0.75) (figure 18) starting with different initial condition of populations. Here $G_2$ also equilibrate at the same value (around 2.5) (figure 19) but at different point of $G_1$. These results of EhSINE1s data indicate about quasispecies model.

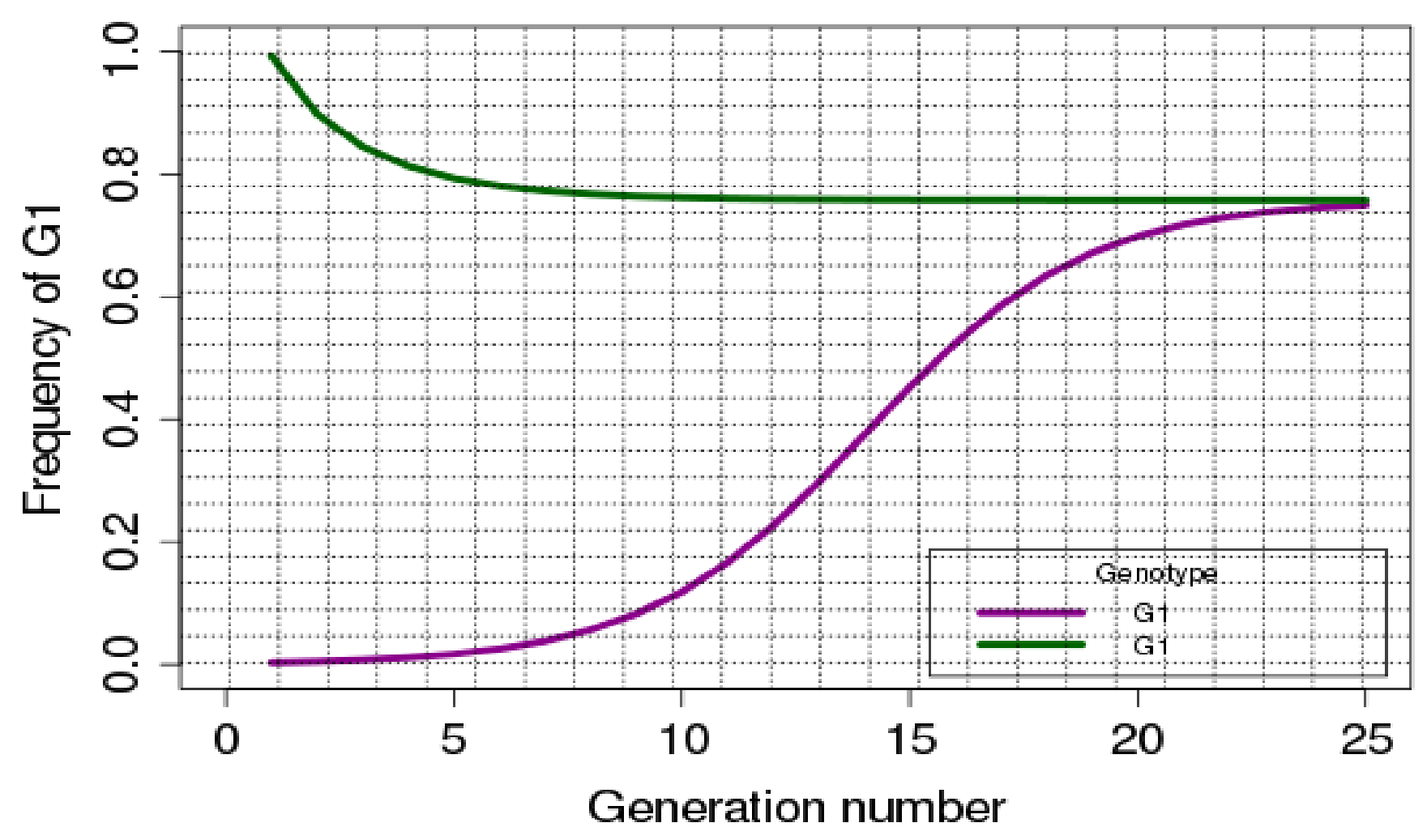

**Figure 18:** Equilibrium approaches of $G_1$ for different initial

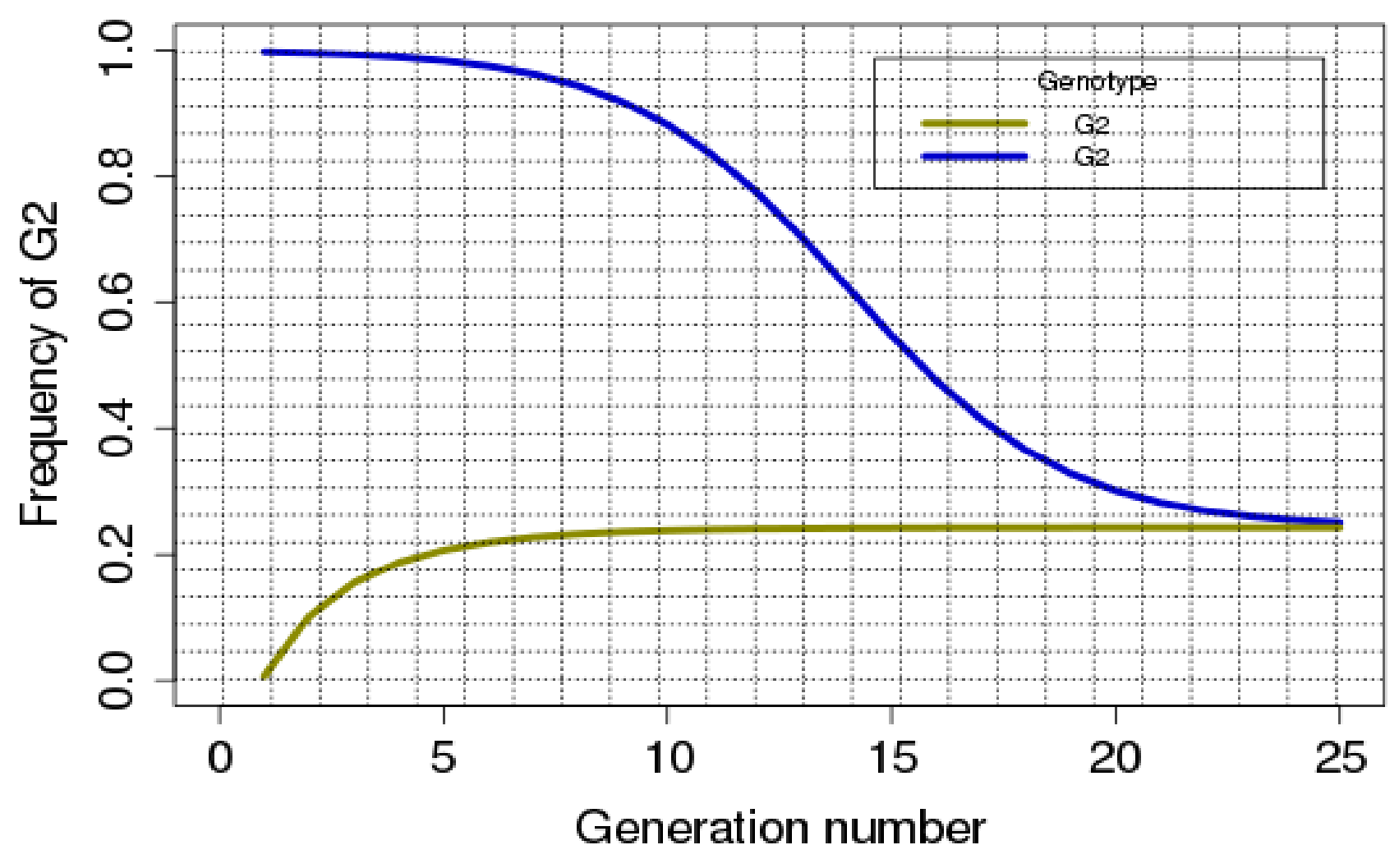

***Figure 19:*** *Equilibrium approaches of $G_2$ with different initial*

### 3.1.6 Results using Quasispecies model with back mutation

Consider a quasispecies model for two genotypes $G_1$ and $G_2$ with back mutation (figure 9). Genotype $G_1$ has mutation rate 0.16 and fitness 0.7 whereas $G_2$ has 0.03 and 0.4. Consider that $G_2$ has back mutation rate $u_3$ = 0.05 (Bull, et al., 2005). The figures (20, 21) are drawn for these parameters. For genotype $G_1$, the upper, dark green curve represents a different starting condition than the lower, dark magenta curve, but they both equilibrate to the same value (around 7.5) after 14 generation (figure 20) and also for genotype $G_2$ both equilibrate to the same value (around 2.5) (see figure 21), but different from the value of G1. Therefore it is easy to say that the back mutation (very small quantity) doesn't make much change in the result of quasispecies model.

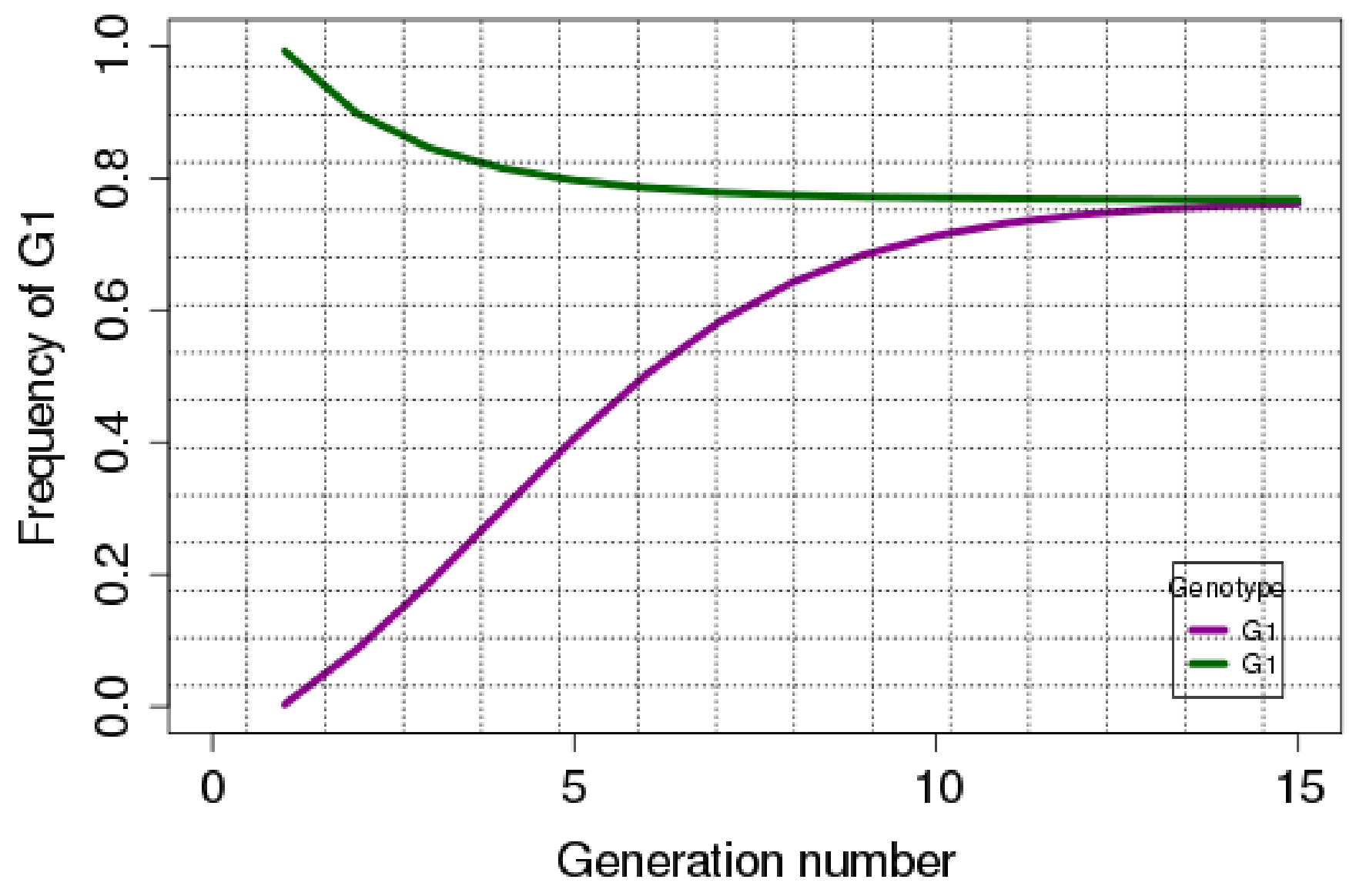

***Figure 20:*** *Equilibrium approaches of* $G_1$ *with different initial conditions in the presence of back mutation*

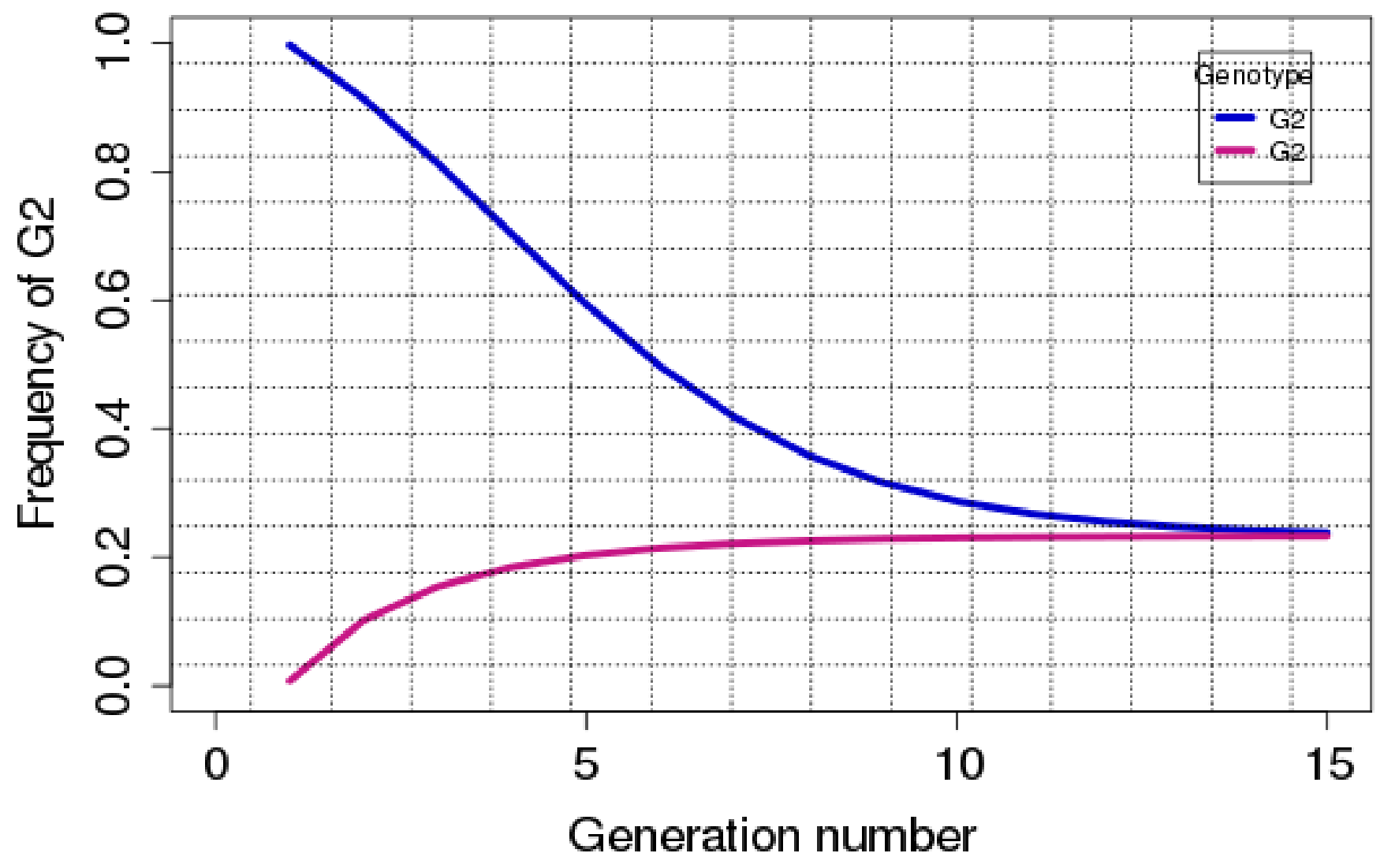

***Figure 21:*** *Equilibrium approaches of* $G_1$ *with different initial conditions in the presence of back mutation*

## 3.2 Analysis of data taken from experiments

EhSINE1 of length 578 bp was grown in vivo (Vijay, et al., 2012) and 24 (including original one) clones were sequenced. We also have analyzed these newly determined 24 sequences of SINE1.

### 3.2.1 Length variation in newly determined EhSINE1s

The lengths of these clones vary from 513 to 578 bp which are shown in figure 22. It can be easily seen that there are four clones which have length 573 bp long and this length is much closer to the length of original EhSINE1. These EhSINE1s are not normally distributed (figure 22).

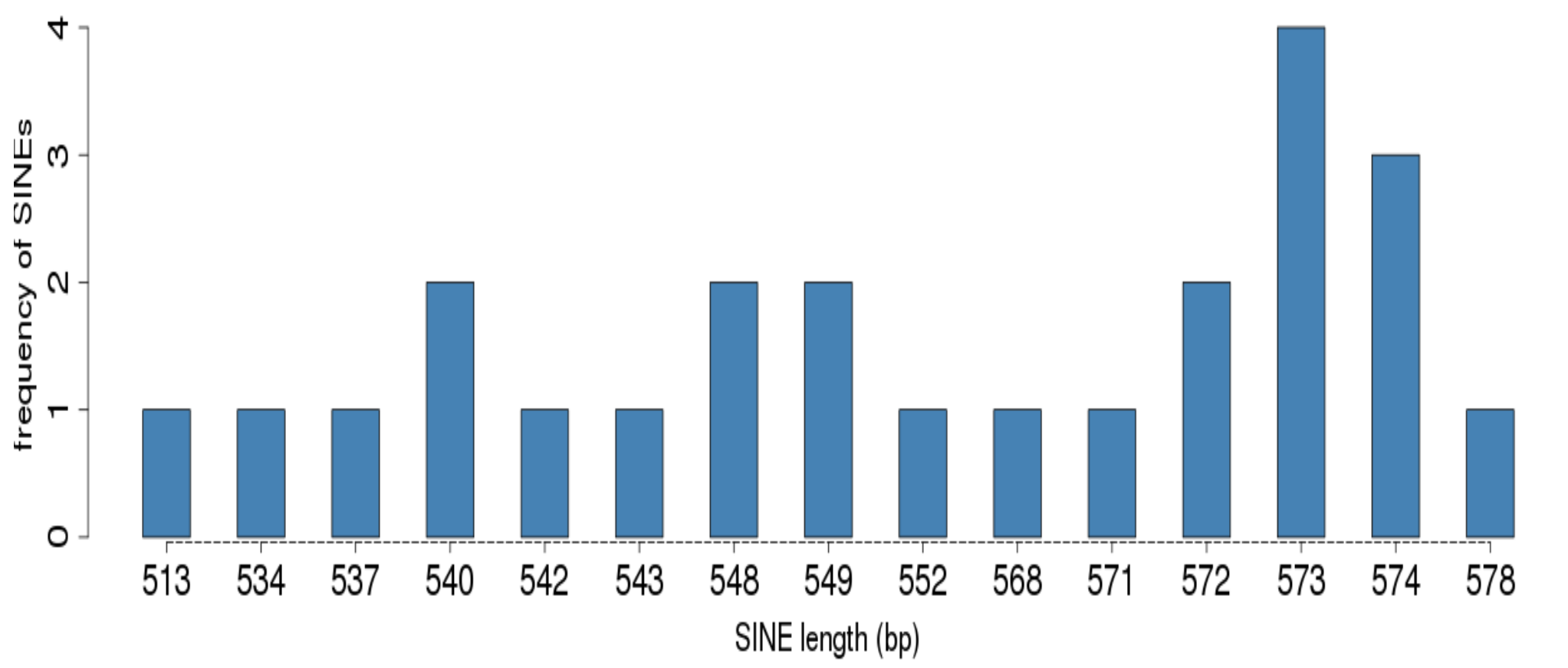

***Figure 22:*** *Distribution of length of 24 clones of EhSINE1*

### 3.2.2 Output of ClutalW for newly determined EhSINE1s

We have used ClustalW for alignment of retrotransposons of *E. histolytica*. Twenty four sequences of dataset II were aligned. A screen shot of alignment of 24 clones in ClustalW is shown in Figure 23. It can be clearly observed from the figure that there are 10 clones which have tag sequence (25 bp long, position 251 to 275) and 8 sequences (from B1 to B8) which do not have tag sequence (position 251 to 275 in dots), and remaining 5 sequences (C1 to C5) contain some part of the original EhSINE1.

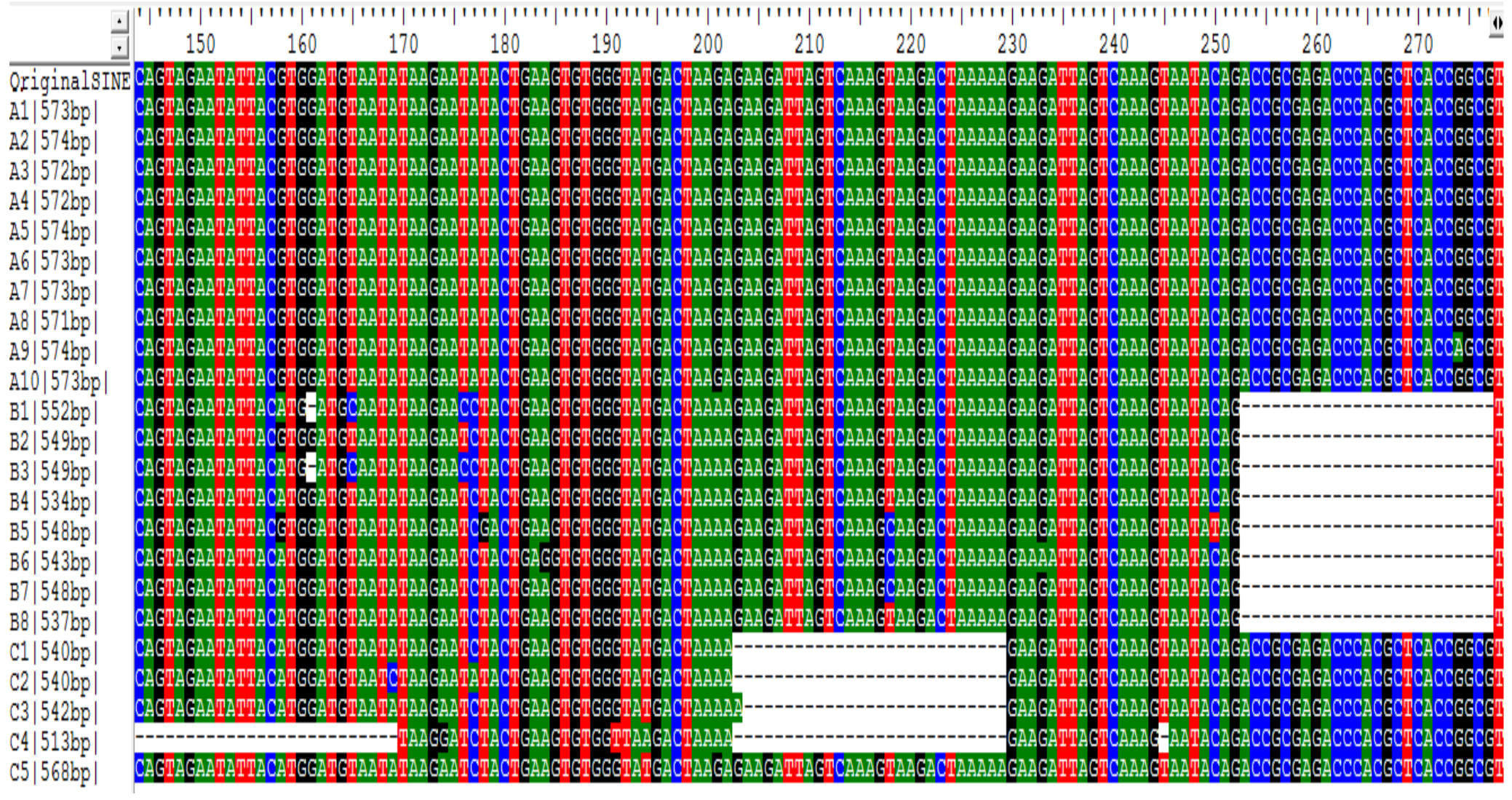

***Figure 23:*** *Multiple sequence alignment of EhSINE1s using ClustalW*

### 3.2.3 Result of phylogenetic tree for newly determined EhSINE1s

The results obtained after phylogenetic analysis is shown below. There are 23 clones (excluding original one) which make 3 groups and we called these groups as genotypes $G_1$, $G_2$ and $G_3$. The clones of genotype $G_1$ are more close to the original EhSINE1.

**Table 6.** Number of groups formed in tree.

| Group Name | $G_1$ | $G_2$ | $G_3$ |
|---|---|---|---|
| Size | 10 | 8 | 5 |

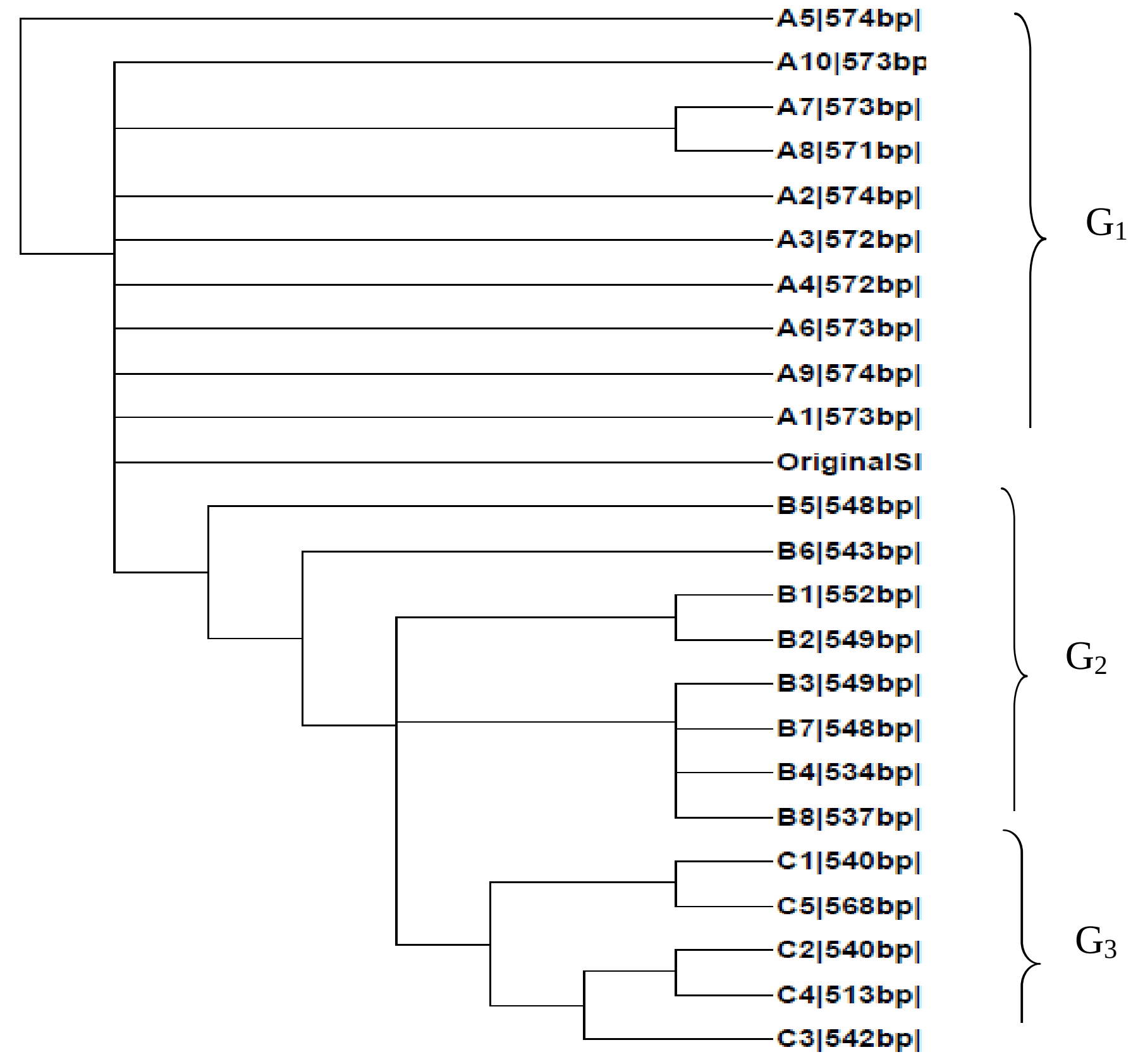

***Figure 24****: Showing phylogenetic tree of newly determined EhSINE1s*

### 3.2.4 Mutation rate of newly determined EhSINE1s

Mutation rates of sequences were calculated in the similar way as discussed above with growth parameter c= 7.5 and mutational target site N = 1/3. Results of mutation rate of these newly determined EhSINE1s are shown in Table 7.

**Table 7.** Mutation rates per base pair ($u_b$) and mutation rates per genome per replication $u_g$ for newly determined EhSINE1s.

| G | Q. Id | S. Id | % Id | A. L. | m | $u_b$ | $u_g$ |
|---|---|---|---|---|---|---|---|
| $G_1$ | OriginalSINE1\|578bp\| | A2\|574bp\| | 99.83 | 575 | 0 | 0 | 0 |
| | | A4\|572bp\| | 99.82 | 570 | 1 | 0.0005 | 0.30 |
| | | A3\|572bp\| | 99.82 | 571 | 0 | 0 | 0 |
| | | A1\|573bp\| | 99.65 | 572 | 1 | 0.0005 | 0.29 |
| | | A9\|574bp\| | 99.48 | 574 | 2 | 0.0010 | 0.59 |
| | | A10\|573bp\| | 99.65 | 569 | 2 | 0.0010 | 0.58 |
| | | A7\|573bp\| | 99.48 | 573 | 2 | 0.0010 | 0.58 |
| | | A6\|573bp\| | 99.48 | 573 | 1 | 0.0005 | 0.29 |
| | | A5\|574bp\| | 99.47 | 571 | 2 | 0.0010 | 0.59 |
| | | A8\|571bp\| | 99.47 | 566 | 3 | 0.0015 | 0.89 |
| $G_2$ | OriginalSINE1\|578bp\| | B8\|537bp\| | 91.01 | 567 | 17 | 0.0093 | 5.35 |
| | | B3\|549bp\| | 90.59 | 574 | 20 | 0.0107 | 6.16 |
| | | B1\|552bp\| | 90.54 | 571 | 21 | 0.0111 | 6.43 |
| | | B6\|543bp\| | 90.48 | 567 | 22 | 0.0118 | 6.85 |
| | | B2\|549bp\| | 90.16 | 569 | 21 | 0.0112 | 6.46 |
| | | B4\|534bp\| | 90.21 | 562 | 22 | 0.0120 | 6.96 |
| | | B5\|548bp\| | 89.56 | 565 | 32 | 0.0171 | 9.87 |
| | | B7\|548bp\| | 97.56 | 246 | 4 | 0.0021 | 1.23 |
| $G_3$ | OriginalSINE1\|578bp\| | C5\|568bp\| | 95.54 | 560 | 17 | 0.0088 | 5.06 |
| | | C3\|542bp\| | 90.53 | 570 | 20 | 0.0108 | 6.24 |
| | | C2\|540bp\| | 90.91 | 561 | 17 | 0.0092 | 5.32 |
| | | C1\|540bp\| | 90.48 | 567 | 21 | 0.0114 | 6.57 |
| | | C4\|513bp\| | 95.24 | 357 | 11 | 0.0063 | 3.62 |
| | | C4\|513bp\| | 90.51 | 137 | 9 | 0.0051 | 2.97 |

The probability that the offsprings of one genotype convert into another genotype is shown in Table 8. Mutation rate of these genotypes of newly determined EhSINE1s are higher than that of the dataset I.

**Table 8.** Mutation rates $u_i$ of genotypes of newly determined EhSINE1s.

| Genotype | Mutation rate $u_i$ |
|---|---|
| G1 | 0.049 |
| G2 | 0.592 |
| G3 | 0.357 |

### 3.2.5 Results of sequence analysis of in vitro generated EhSINE1s using quasispecies model without back mutation

Consider a quasispecies model for two genotypes (figure 7) of newly generated EhSINE1s where first genotype (population size $p_i$ = 3000; i=1,2,3) is fully dominant to the second one (population size $p_j$ = 20; j=1,2,3). Now consider again a model for different condition i.e. second dominant ($p_j$=2000; j=1,2,3) to the first one ($p_i$ = 15; i=1,2,3). Using this population size, results for different values of fitness, using quasispecies model, are described.

**Result for fitness 1.5 and 1:**

Suppose a model of two genotype $G_3$ and $G_2$ with mutation rates $u_3$ = 0.357 and $u_2$ = 0.592 (from Table 8), and fitness $f_3$ = 1.5 and $f_2$ = 1 (Bull, et al., 2005). Here $G_3$ is the first one and $G_2$ is the second. The figures (25, 26) are drawn for these parameters. For genotype $G_2$, the upper, dark red curve represents a different starting condition than the lower, dark violet curve, but they both equilibrate to the same value (around 6.0) after 10 generation (figure 25) and for genetype $G_3$, both equilibrate to the same value but different from the value of $G_2$ (figure 26). Here it can be seen that the population equilibrate very soon in comparison to previous result (dataset I), it is because a combination of high mutation rate and fitness.

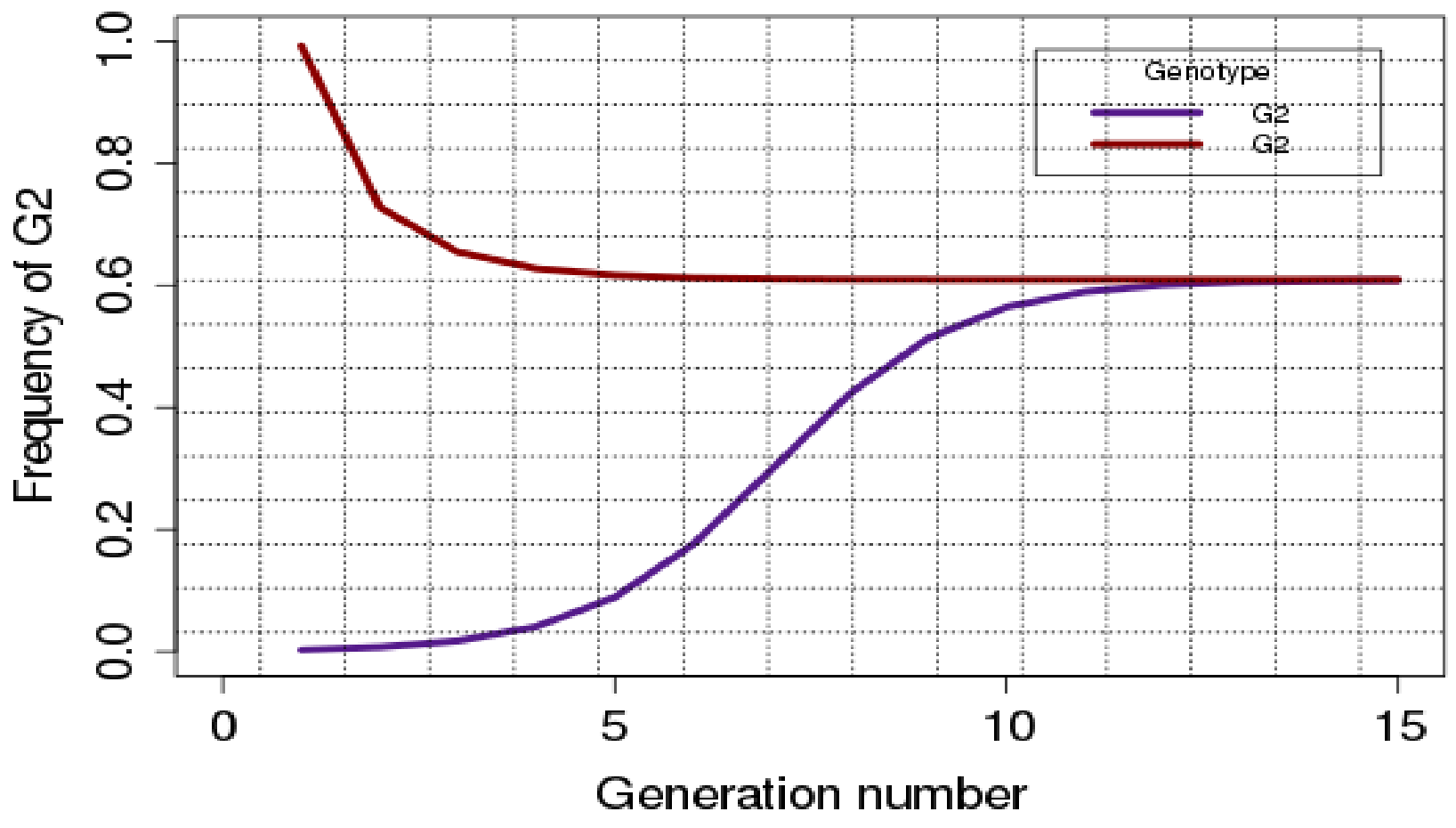

***Figure 25:*** *Equilibrium approaches of genotype $G_2$ of newly determined EhSINE1s with different initial conditions.*

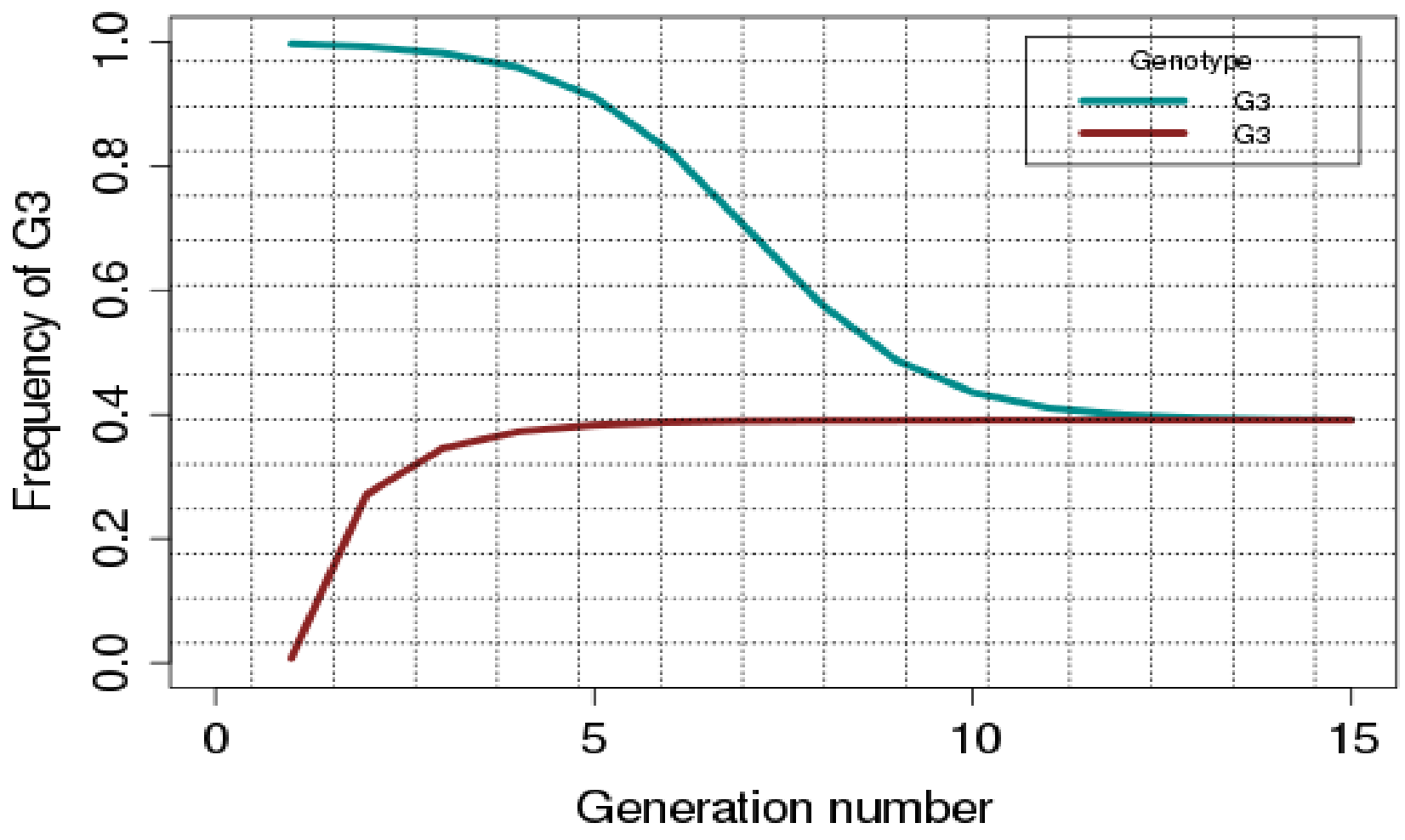

***Figure 26:*** *Equilibrium approaches of genotype $G_3$ of newly determined EhSINE1s with different initial conditions*

**Result for fitness 0.5 and 1:**

Consider same model for genotype $G_1$ and $G_2$ with fitness $f_1 = 0.5$ and $f_2 = 1$ and mutation rate $u_1 = 0.049$ and $u_2 = 0.592$. Using these values, the results obtained are shown below (figures 27, 28). The population of $G_1$ for two different conditions equilibrates after a long generation (generation 60) in comparison to previous results, it is because of combination of very low mutation rate and fitness of genotype $G_1$ and very high mutation rate and fitness of genotype $G_2$.

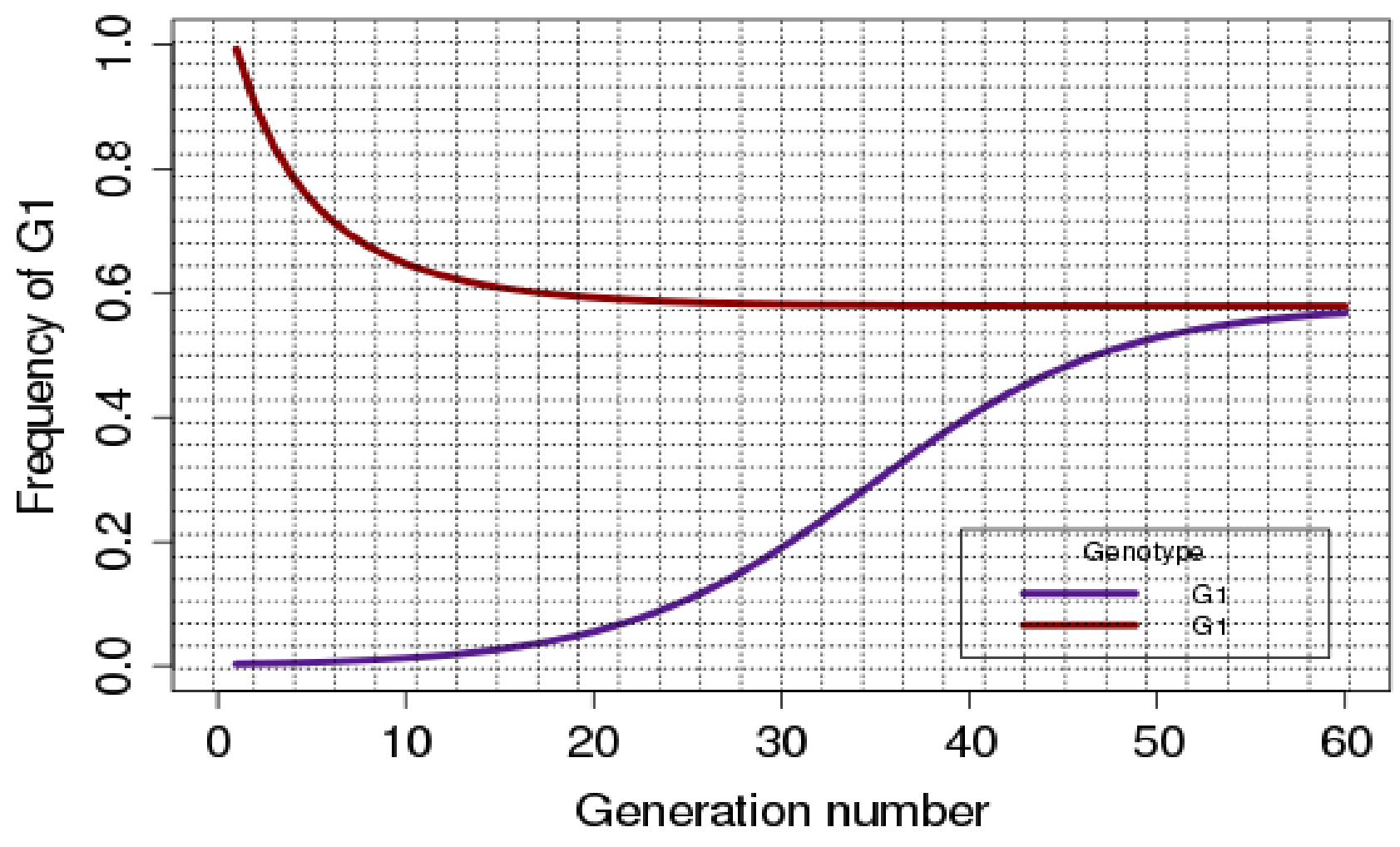

***Figure 27:*** *Equilibrium approaches of genotype $G_1$ of newly determined EhSINE1s with different initial conditions*

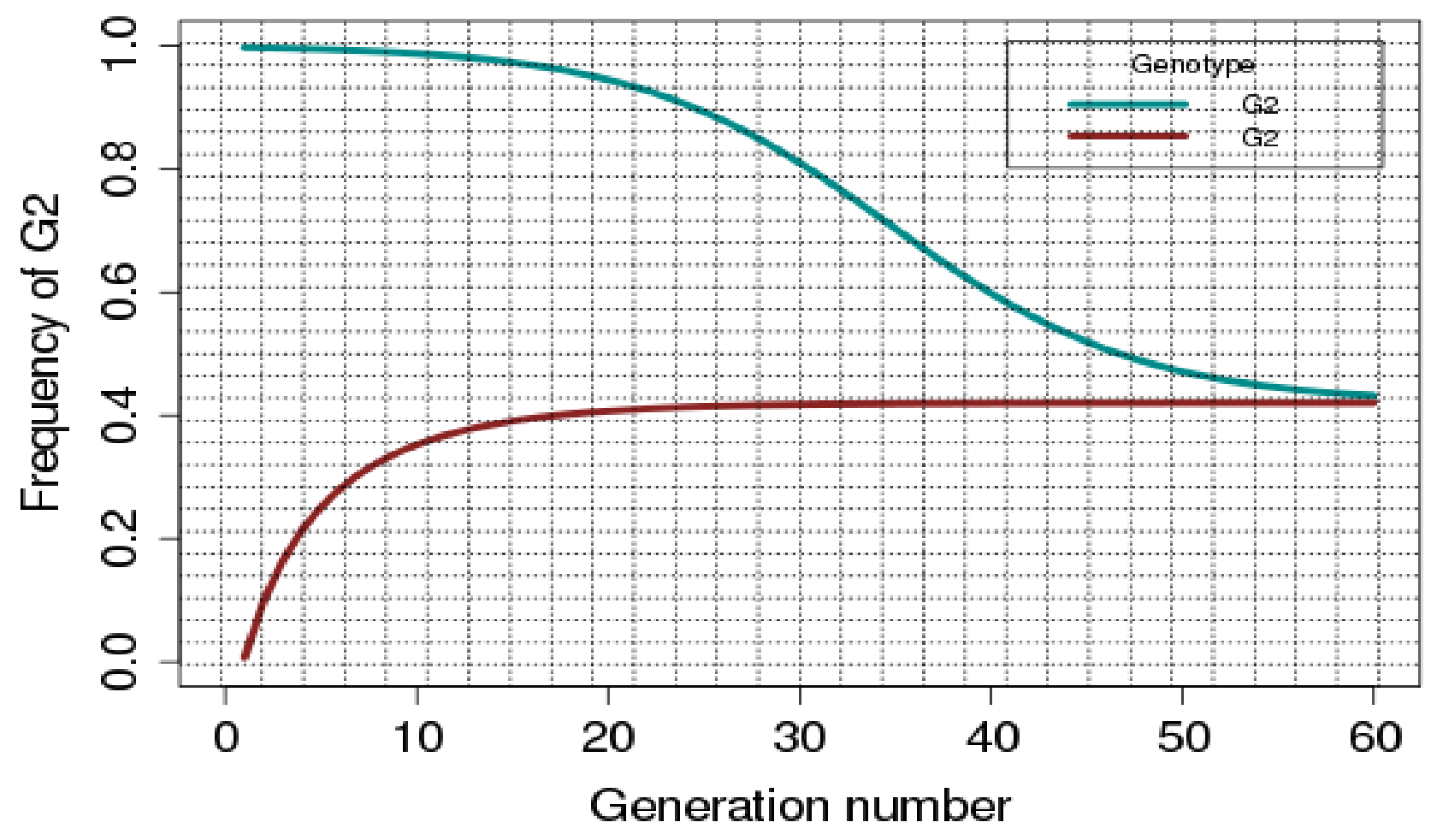

***Figure 28:*** *Equilibrium approaches of genotype $G_2$ of newly determined EhSINE1s with different initial conditions*

# Chapter 4

# Summary and Conclusion

From our study of Non-long terminal repeat retrotransposons (LINEs and their non-autonomous partners SINEs) of Entamoeba histolytica we can conclude that an active EhSINE can generate very similar copies of itself by retrotransposition. Due to this reason it increases mutations which give the result of sequence polymorphism. We have concluded that the mutation rate of SINE is very high. This high mutation rate provides an idea for the existence of SINEs, which may affect the genetic analysis of EhSINE1 ancestries, and calculation of phylogenetic distances.

From the study of quasispecies model for EhSINE1s it can be concluded that for any combination of mutation rates and fitnesses, a population will develop until it arrives at a unique equilibrium distribution of $G_1$ and $G_2$. The evolution of two populations in which one that initially consisted entirely of $G_1$ and another that was generated due to changes ($G_2$), finally both arrived at the same equilibrium proportions. This happens because of the same mutation rates and fitnesses in both cases. The equilibrium distribution will change with changing mutation rate or fitness. Therefore, the correspondent figures for different equilibria points would have the curves converging to a distinct value on y axis. This equilibrium point is the mutation selection balance where mutation continually produces $G_2$ from $G_1$ and natural selection continually disgorges $G_2$ in favor of $G_1$ and it refer as a quasispecies.
Therefore we can conclude that the changes in the sequences of the retrotransposons (LINEs and SINEs) of *E. histolytica* follow the quasispecies model.